\documentclass[12pt,preprint]{aastex}

\shorttitle{Very massive stars outside the NGC3603 cluster center}
\shortauthors{A. Roman-Lopes, G. A. P. Franco \& D. Sanmartin}

\begin{document}


%%\title{SOAR optical blue and near-infrared spectra of newly discovered massive stars 
%%in Galactic star clusters: I - New OIf* and OIf*/WN stars found in the outskirt of NGC3603}

\title{SOAR optical and near-infrared spectroscopic survey of newly discovered massive stars in the periphery of Galactic Massive star clusters I - NGC3603}


\author{A. Roman-Lopes\altaffilmark{}}
\affil{Department of Physics and Astronomy -- Universidad de La Serena, Cisternas 1200, La Serena, Chile}
\email{aroman@userena.cl}

\author{G. A. P. Franco\altaffilmark{}}
\affil{Departamento de F\'isica -- ICEx -- UFMG, Caixa Postal 702, 30.123-970, Belo Horizonte, MG, Brazil}

%%\and

\author{D. Sanmartim\altaffilmark{}}
\affil{Southern Astrophysical Research Telescope (SOAR), Chile}


%%\altaffiltext{1}{Visiting Astronomer, Cerro Tololo Inter-American Observatory.
%%CTIO is operated by AURA, Inc.\ under contract to the National Science
%%Foundation.}
%%\altaffiltext{2}{Society of Fellows, Harvard University.}
%%\altaffiltext{3}{present address: Center for Astrophysics,
%%    60 Garden Street, Cambridge, MA 02138}
%%\altaffiltext{4}{Visiting Programmer, Space Telescope Science Institute}
%%\altaffiltext{5}{Patron, Alonso's Bar and Grill}


\begin{abstract}

In this work, we present the results of a spectroscopic study of very massive stars found outside the center of the massive stellar cluster NGC3603.
From the analysis of the associated SOAR spectroscopic data and related optical-NIR photometry, we confirm the existence of several very massive stars in the 
periphery of NGC 3603. The first group of objects (MTT58, WR42e and RFS7) is compound by three new Galactic exemplars of the OIf*/WN type, all of them with probable initial masses 
well above 100 M$_{\odot}$ and estimated ages of about 1 Myrs. Based on our Goodman blue-optical spectrum of MTT68, we can confirm 
the previous finding in the NIR of the only other Galactic exemplar (besides HD93129A) of the O2If* type known to date. Based on its position relative to a 
set of theoretical isochrons in a Hertzprung-Russel diagram, we concluded that the new O2If* star could be one of the most massive (150 M$_{\odot}$) and 
luminous (M$_V$=-7.3) O-star in the Galaxy. 
Also, another remarkable result is the discovery of a new O2V star (MTT31) that is the first exemplar of the class so far identified in the Milk Way. From 
its position in the Hertzprung-Russel diagram it is found that this new star probably had an initial mass of 80 M$_{\odot}$, as well as an absolute 
magnitude M$_V$=-6.0 corresponding to a luminosity similar to other known O2{\sc v} stars in the LMC.
Finally, we also communicate the discovery of a new Galactic O3.5If* star (RFS8) which case is quite intriguing. Indeed, It is located far to the south of 
the NGC 3603 center, in apparent isolation at a large radial projected linear distance of $\sim$ 62 pc. Its derived luminosity is similar to that of 
the other O3.5If* (Sh18) found in the NGC 3603's innermost region, and the fact that a such high mass star is observed far isolated in the field led 
us to speculate that perhaps it could have been expelled from the innermost parts of the complex by a close fly-by dynamical encounter with a very 
massive hard binary system.

\end{abstract}





\section{Introduction}

Very Massive Stars (VMS) \citep{vink15} are expected to be found in the core of their host clusters, generally forming binary or multiple stellar systems. Indeed, it is well accepted that the majority of massive stars are formed in clusters
with the O3 stars being considered for a long time, the most massive hydrogen core burning stellar type. However, the situation has changed in the last few 
decades, as we now know that some hydrogen-rich nitrogen sequence Wolf-Rayet (WR) stars are in reality extremely massive and luminous main-sequence (MS) stars, 
which because the proximity to the Eddington limit, mimic the spectral appearance of classical WR stars showing an emission-line spectrum even at the beginning 
of their main-sequence evolution \citep{dekoter97,schnurr08,smith08,crow10}.

Several observational studies on stellar masses of very massive binary systems, indicate that such massive stellar objects belong to the OIf*/WNH and WNH 
spectral types \citep{smith08,crow10,crow11}, making them the most massive MS stars known in the local universe. Evidence supporting this assumption come from 
systematic studies of binaries made by \citet{rauw96} for WR22 (minimum masses of 71.7$\pm$2.4 M$_\odot$ + 25.7$\pm$0.8 M$_\odot$), \citet{rauw04} and \citet{bona04} 
for WR20a (two O3If*/WN6 stars with absolute masses of 83 M$_\odot$ and 82 M$_\odot$), \citet{nieme08} for WR21a (a WN6ha star with minimum mass of 
87 M$_\odot$ + O-type secondary with a minimum mass of 53 M$_\odot$), \citet{schnurr08} for NGC3603-A1 
(minimum masses of 116$\pm$31 M$_\odot$ + 89$\pm$16 M$_\odot$), \citet{schnurr09} for R145 
(minimum masses of 116$\pm$33 M$_\odot$ + 48$\pm$20 M$_\odot$). Finally, \citet{crow10} found from their spectroscopic re-analyses of OIf*/WN and WNH stars 
in the cores of NGC 3603 and R136 in the Large Magelanic Cloud (LMC), that some stars there might had initial masses in the range of 105-170 M$_\odot$ and 
165-320 M$_\odot$, respectively.

%%Table 1
\begin{table}
\begin{center}
\caption{Number of O2If*, OIf*/WN and WNh stars previously known in the Wd2 and NGC3603 fields, found at R$_c$ distances in intervals R$_1$=(0 $<$ R$_c$ $<$ r), R$_2$=(r $<$ R$_c$ $<$ r$_t$) and R$_3$=R$_c$ $>$ r$_t$. The parameters r$_c$, r and r$_t$ are respectively the observed cluster core radius, the cluster radius and the cluster tidal radius of NGC 3603 and Wd2 derived by \citet{hur14} and \citet{sung04}.\label{tbl-1}}
\begin{tabular}{ccccccc}
\tableline\\
Cluster & r$_c$ ($\arcmin$)& r ($\arcmin$)& r$_t$ ($\arcmin$)& [ 0 - r] & [r - r$_t$] & [$>$ r$_t$] \\
\tableline
Wd2 & 0.2 & 1.78 & 4.8 & 1 & 1 & 3\\
NGC3603 & 0.05 & 2.0 & 14.6 & 7 & 1 & 0 \\
\tableline
\end{tabular}
\end{center}
\end{table}
%%%End Table

Surprisingly, some VMS are found in isolation in the field. As examples we can mention the cases of WR21a (a binary with minimum masses 
87 M$_\odot$ + 53 M$_\odot$ - \citet{nieme08}) in Westerlund 2 (Wd2), WR42e (an O3If*/WN6 with estimated initial mass above 100 M$_\odot$ - \citet{roman12,gvara13}) 
in NGC 3603, and VFTS 682 in the Large Magellanic Cloud (LMC) whose blue-optical spectra looks very similar to that of R136a3 (one of the WNH stars found in the 
centre of R136), with an inferred initial mass of $\sim$ 150-200 M$_\odot$ - \citet{besten11,vink15}). In this sense, there is a growing number of VMS that are 
seen outside the cluster cores, sometimes found far away from their supposed parental clusters, perhaps forming massive stellar halos like the 
one seen in the field of the 30 Dor starburst region in the LMC \citep{walb94,walb14}. In the Milk Way, this phenomenon seem to occur in the periphery of massive 
stellar clusters like Wd2 and NGC 3603, in which some of the most massive stellar members are found far from the cluster centers. For example, in Wd2 
besides WR20a there are WR20b (WN6ha - \citet{moffat91,shara91}) and WR21a (WN6 + early-O) \citep{nieme08}, which are found in isolation at angular 
radial cluster center distances (R$_c$) of 0.6\arcmin, 3.7\arcmin, and 16\arcmin, respectively, plus two O2If*/WN6 stars, WR20aa and WR20c \citep{roman11}, 
placed at large R$_c$ values of 15.7\arcmin, and 25\arcmin, respectively. On the other hand, in the NGC3603's field there are three 
exemplars: MTT58 (O2If*/WN6), MTT68 (O2If*) and WR42e (O2If*/WN6) at R$_c$ values of 1.0\arcmin, 1.4\arcmin, and 3\arcmin, 
respectively \citep{roman12,roman13a,roman13b}. 

For comparison, in Table 1 it is shown the number of VMS found in the direction of Wd2 and NGC 3603, tabulated for R$_c$ values in the ranges 
0 $<$ R$_c$ $<$ \textit{r}, \textit{r} $<$ R$_c$ $<$ $r_t$ and R$_c$ $>$ $r_t$, with \textit{r} and $r_t$ being the observed cluster radius and the 
cluster tidal radius, respectively, taken from \citet{hur14} and \citet{sung04}.
We can see that the number of O2If*, OIf*/WN and WNH objects found inside the circular area limited by the cluster radius \textit{r} is higher 
in NGC 3603 than in Wd2, perhaps reflecting differences in the total mass of each cluster and/or in their dynamical evolutionary stages. On the other 
hand, no such objects are found beyond the area projected by the respective tidal radius, a behavior that led us to wonder if there would be some other 
VMS still to be discovered in the outskirt of the NGC3603 complex. 

In this work we performed a search for VMS candidates placed beyond the center of the massive stellar cluster NGC3603, which is known to be one of the most 
massive, dense and rich Galactic star-forming region \citep{melnick89,moffat02,sung04}, and believed to be a scaled version of the star-burst R136 cluster in the 
Large Magellanic Cloud (LMC). It has several massive stars in its core, many of them  apparently showing initial masses up to 100-170 M$_\odot$ \citep{crow10}.
Based on near-infrared (NIR) colour and magnitude selection criteria applied to objects found in the 2MASS point source 
catalogue - 2MASS PSC - \citet{skrut06}, the chosen stars were surveyed through a SOAR NIR spectroscopic survey aimed to confirm their possible massive nature. 
The earliest stars selected from our analysis of the respective NIR spectra were then re-studied and spectroscopically re-classified using new SOAR-Goodman 
blue-optical data, resulting on the confirmation of the existence of several new massive stars in the NGC3603 field.

%%Table 2

\begin{deluxetable}{ccccccccccc}
\tabletypesize{\scriptsize}
\rotate
\tablecaption{List of stars observed with Osiris and Goodman with the SOAR telescope. Column 1 is the assigned ID, columns 2 and 3 their coordinates (J2000), 
columns 4 and 5 the B-, V-band photometry taken from \citet{sung04,zach04,zach13}, columns 6-8 the J-, H- and K$_S$-band photometry taken from the 2MASS PSC. 
Column 9 the corresponding NGC3603 radial angular and projected linear distance (parsecs - assuming a heliocentric distance of 7.6 kpc \citep{crow10}) 
center distances, and column 10 the observed wavelength ranges. Finally, in column 11 we list the IDs of the sources found in the literature, as well as 
the objects with (when available) X-ray band measurements. The previously assigned IDs for some of the sources listed here 
(like RFS1, RFS2, RFS3, and RFS4) are from the work of \citet{melnick89}, while WR42e was previously named by \citet{roman13b}. 
Also the informations on the X-ray source counterparts are from the catalog of \citet{romano08} (for RFS1 to RFS5), and in case of RFS10, 
from the XMM-Newton Serendipitous Source catalog \citep{xmm13}.}
\tablewidth{0pt}
\tablehead{
\colhead{Source} & \colhead{RA(J2000)} & \colhead{Dec(J2000)} & \colhead{B} & \colhead{V} & 
\colhead{J} & \colhead{H} & \colhead{K$_S$} & \colhead{r (arcmin - pc)} & 
\colhead {OSIRIS Data} & \colhead{comments}
}
\startdata

RFS1 & 11:15:06.68 & -61:16:33.3 & 15.29 & 14.07 & 10.67 & 10.17 & 9.86 & 0.7 - 1.5 & Optical+NIR & MTT31$^{(1)}$  (O4V-O5V and X-Ray) \\
RFS2 & 11:15:07.58 & -61:16:54.6 & 16.14 & 14.76 & 10.47 & 9.68 & 9.24 & 1.0 - 2.1 & Optical+NIR & MTT58 (X-Ray)\\
RFS3 & 11:14:59.48 & -61:14:33.8 & 16.31 & 14.72 & 9.98 & 9.17 & 8.74 & 1.4 - 3.0 & Optical+NIR & MTT68 (X-Ray)\\
RFS4 & 11:15:21.32 & -61:15:04.3 & 16.31 & 14.74 & 10.60 & 9.99 & 9.60 & 1.8 - 3.7 & Optical+NIR & MTT71 (X-Ray) \\
RFS5 & 11:14:45.50 & -61:15:00.1 & 16.05 & 14.53 & 10.18 & 9.47 & 9.04 & 3.0 - 6.4 & Optical+NIR & WR42e (X-Ray) \\
RFS6 & 11:14:53.55 & -61:24:22.8 & 16.02 & 14.52 & 10.64 & 10.06 & 9.73 & 8.5 - 18 & Optical+NIR & X-Ray source \\
RFS7 & 11:15:15.36 & -60:51:17.6 & 14.17 & 12.89 & 9.85 & 9.39 & 9.12 & 25 - 53 & Optical+NIR & --- \\
RFS8 & 11:16:12.62 & -61:43:54.2 & 15.31 & 14.68 & 10.61 & 9.92 & 9.48 & 29 - 62 & Optical+NIR & --- \\
RFS9 & 11:12:53.35 & -60:50:45.2 & 14.98 & 13.62 & 10.55 & 10.06 & 9.83 & 30 - 64 & Optical+NIR & --- \\
RFS10 & 11:19:55.12 & -61:16:03.7 & 15.21 & 13.86 & 10.54 & 10.03 & 9.66 & 35 - 74 & Optical & X-Ray source \\

\enddata
\tablecomments{(1) Previous spectral type for MTT31 (RFS1) as presented by \citet{moffat02}}.  
%A portion is 
%shown here for guidance regarding its form and content.}
%\tablenotetext{a}{Sample footnote for table~\ref{tbl-1} that was generated
%with the deluxetable environment}
%\tablenotetext{a}{The column SpType presents the spectral types derived in previous studies using only NIR spectra}
\end{deluxetable}

\section{Near-IR magnitude and color-ratio selection criteria}

The stars of this work were selected from the study of the near-infrared (NIR) magnitudes and colors of sources in the 2MASS PSC found on the annular sky area 
(centered on the NGC3603's coordinates) of internal and external radius 0.5\arcmin, and 35\arcmin, respectively, and presenting K$_S$-band magnitudes in the 
range 8.5$<$K$_S$$<$10 and magnitude errors $<$ 0.1. For the sources matching the mentioned spatial and magnitude criteria, we kept those showing colors and 
color-ratios in the range 0.5$<$(J-K$_S$)$<$1.5 and 1.5$<$[(J-H)/(H-K$_S$)]$<$2.0, respectively, which in turn were defined based on the fact that 
massive stars are known to emit a lot of excess of radiation in the infrared and radio domains, mainly due to the free-free emission generated by their 
powerful stellar winds (for more on this see \citet{lamers99} and references therein). In this sense, the K$_S$-band magnitude range and color criteria 
were tuned based on those of WR42e (O2If*/WN6), MTT58 (O2If*/WN6) and MTT68 (O2If*), with the K$_S$-band magnitude limit (K$_S$=10) being fixed considering 
the maximum integration time previously assigned per source (about 30-35 minutes) with OSIRIS at SOAR, taking into account a minimum S/N = 100 in the same 
spectral band. This empirical limit possibly results in a bias towards the most luminous (and potentially massive) objects, what in principle is not an issue 
as our main goal is to do it for the most promising candidates. 
In one hand, by using the mentioned magnitude limits we can probably be missing several mid- to late-O type stars. Indeed, as an example of this behavior we 
can mention the case  of the newly discovered O6{\sc v} runaway star 2MASS J11171292-6120085 \citep{gvara13} that is found at 15.6$\arcmin$ from the NGC3603's 
center, which is not selected by our methodology because its K$_S$ magnitude (10.92) is well above the assumed upper selection limit. On the other hand, we are 
also probably missing a certain number of massive stars positioned in the innermost regions of the complex, e.g. those with radial distances $r<$0.5\arcmin.

%-----------------------------------------------------------------------------------------------------------------------------------------------------------
%Figure1
   \begin{figure}
 %   \vspace{0pt}
    %\hspace{25pt}
  \centering
   \includegraphics[width=16.5 cm]{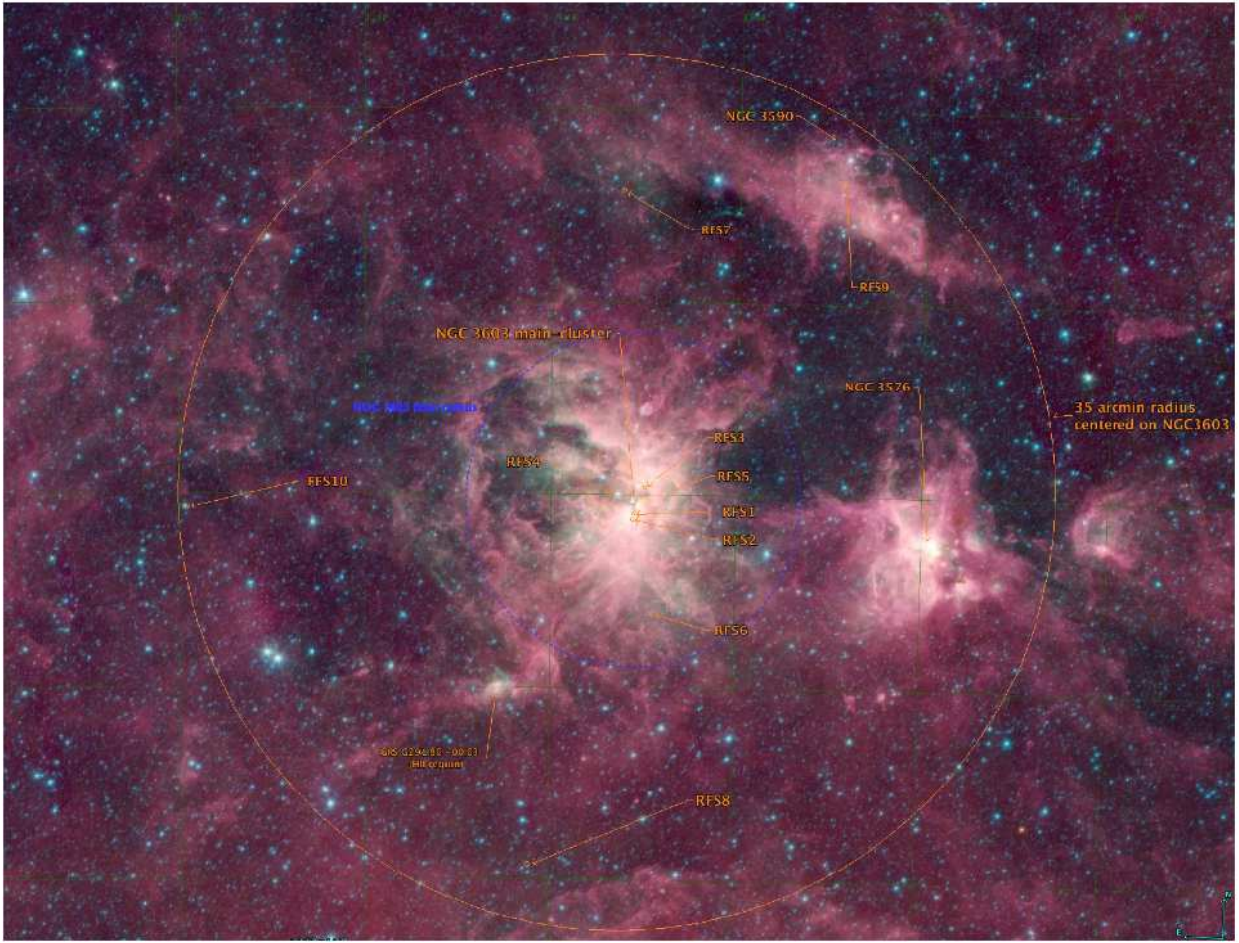}
     \caption{A colorized Spitzer image (blue-3.6$\mu$m, green-4.5$\mu$m, red-8.0$\mu$m) in the direction of the NGC 3603 complex, in which we indicate the position of each source in our sample (shown by labels and arrows), as well as their relative position to NGC 3603, NGC 3576 and NGC 3590. North is to the top and east to the left. The approximate tidal radius (about 15 arcmin) area is indicated by the blue dotted line.}
        \label{FigVibStab}
  \end{figure}   
%-----------------------------------------------------------------------------------------------------------------------------------------------------------

From the application of the above criteria, about thirty 2MASS point sources were selected. Two of them are the known WR stars WR42c and WR42d 
(both of the WN5 type \citep{ross15}), placed at NGC3603 radial distances of 15$\arcmin$ and 6$\arcmin$, respectively.
On the other hand, in the innermost regions there are two known O-stars, SHER 22 (O2-O3 III) and SHER 23 (OC9.7 Ia), both found at less than 
0.5$\arcmin$ from the NGC3603 center.\
The ten most extreme objects of the remaining selected sources are the subject of this paper. Their coordinates and photometric parameters, 
together with the assigned spectral types and NGC3603 radial distances, are listed in Table 2. In Figure 1 we present a colorized Spitzer 
image centered on the NGC 3603 complex, in which we indicate the position of each selected source (shown by labels and arrows), as well as 
their relative position to NGC 3603, NGC 3576 and NGC 3590 star formation complexes.

%In the next section we present the results of our near-infrared and optical spectroscopic surveys.

%%Table 3

\begin{deluxetable}{cccccccc}
\tabletypesize{\scriptsize}
%\rotate
\tablecaption{Journal of the SOAR spectroscopic data used in this work. \label{tbl-1}}
\tablewidth{0pt}
\tablehead{
\colhead{Night} & \colhead{UT} & \colhead{Seeing (\arcsec)} & \colhead{Instrument} & \colhead{Mode} &
\colhead{Slit} & \colhead{Resolving Power} & \colhead{Coverage}
}
\startdata

09-05-11 & 04:10:00-04:22:00 & 0.8-1.0 & OSIRIS & XD - f/3 & 1$\arcsec$ & 1000 & 1.25-2.35$\mu$m \\
18-12-11 & 07:22:00-09:13:00 & 0.8-1.0 & OSIRIS & XD - f/3 & 1$\arcsec$ & 1000 & 1.25-2.35$\mu$m \\
31-01-12 & 06:06:00-08:25:00 & 0.8-1.0 & OSIRIS & XD - f/3 & 1$\arcsec$ & 1000 & 1.25-2.35$\mu$m \\
29-12-12 & 06:02:00-06:58:00 & 0.6-0.8 & OSIRIS & XD - f/3 & 1$\arcsec$ & 1000 & 1.25-2.35$\mu$m \\
26-01-13 & 02:55:00-07:06:00 & 1.0-1.5 & OSIRIS & XD - f/3 & 1$\arcsec$ & 1000 & 1.25-2.35$\mu$m \\
28-02-13 & 08:22:00-09:09:00 & 1.0-1.2 & Goodman & GG385 - 600 l/mm & 1.03$\arcsec$ & 1800 & 0.45-0.67$\mu$m \\
30-03-13 & 07:05:00-07:46:00 & 0.6-0.8 & OSIRIS & XD - f/3 & 1$\arcsec$ & 1000 & 1.25-2.35$\mu$m \\
08-03-15 & 04:54:00-08:05:00 & 1.0-1.5 & Goodman & 930 - m2 & 1.03$\arcsec$ & 2100 & 0.39-0.55$\mu$m \\
28-03-15 & 02:15:00-05:16:00 & 1.0-1.5 & Goodman & 930 - m2 & 1.03$\arcsec$ & 2100 & 0.39-0.55$\mu$m \\
29-06-15 & 22:48:00-03:08:00 & 0.8-1.3 & Goodman & 930 - m2 & 1.03$\arcsec$ & 2100 & 0.39-0.55$\mu$m \\
\enddata
%\tablecomments{Table \ref{tbl-1} is published in its entirety in the 
%electronic edition of the {\it Astrophysical Journal}.  A portion is 
%shown here for guidance regarding its form and content.}
%\tablenotetext{a}{Sample footnote for table~\ref{tbl-1} that was generated
%with the deluxetable environment}
%\tablenotetext{b}{Another sample footnote for table~\ref{tbl-1}}
\end{deluxetable}

\section{Spectroscopic data}

In this section we present details on the observations and data reduction process of the Ohio State Infrared Imager and Spectrometer (OSIRIS) NIR 
and Goodman optical spectroscopic data taken at the Southern Astrophysical Research (SOAR) Telescope for sources listed in Table 2. All but one of them (RFS10), 
were observed with both instruments. Our strategy was to observe the sources selected from the photometric criteria, first through the NIR window using OSIRIS, 
because many of the science targets are relatively faint at the U- and V-bands, mainly due to the large heliocentric distances combined with heavier interstellar 
absorption. The typical integration time necessary to get a S/N$\approx$100 (at $\approx$4800\AA) for a science target with V=14-15 (using Goodman at SOAR in a 
gray time night) is about 1.0-1.5 hours, while if we consider the K$_S$-band magnitude range as defined in Section 2.1, the necessary average exposure time with 
OSIRIS at SOAR (to get a S/N$\approx$100 in the K$_S$-band) drops to no more than 15 to 20 minutes.

%\clearpage
%--------------------------------------------------------K$_S$---------------------------------------------------------------------------------------------------
%Figure2
   \begin{figure}
 %   \vspace{0pt}
    %\hspace{25pt}
  \centering
   \includegraphics[width=16.5 cm]{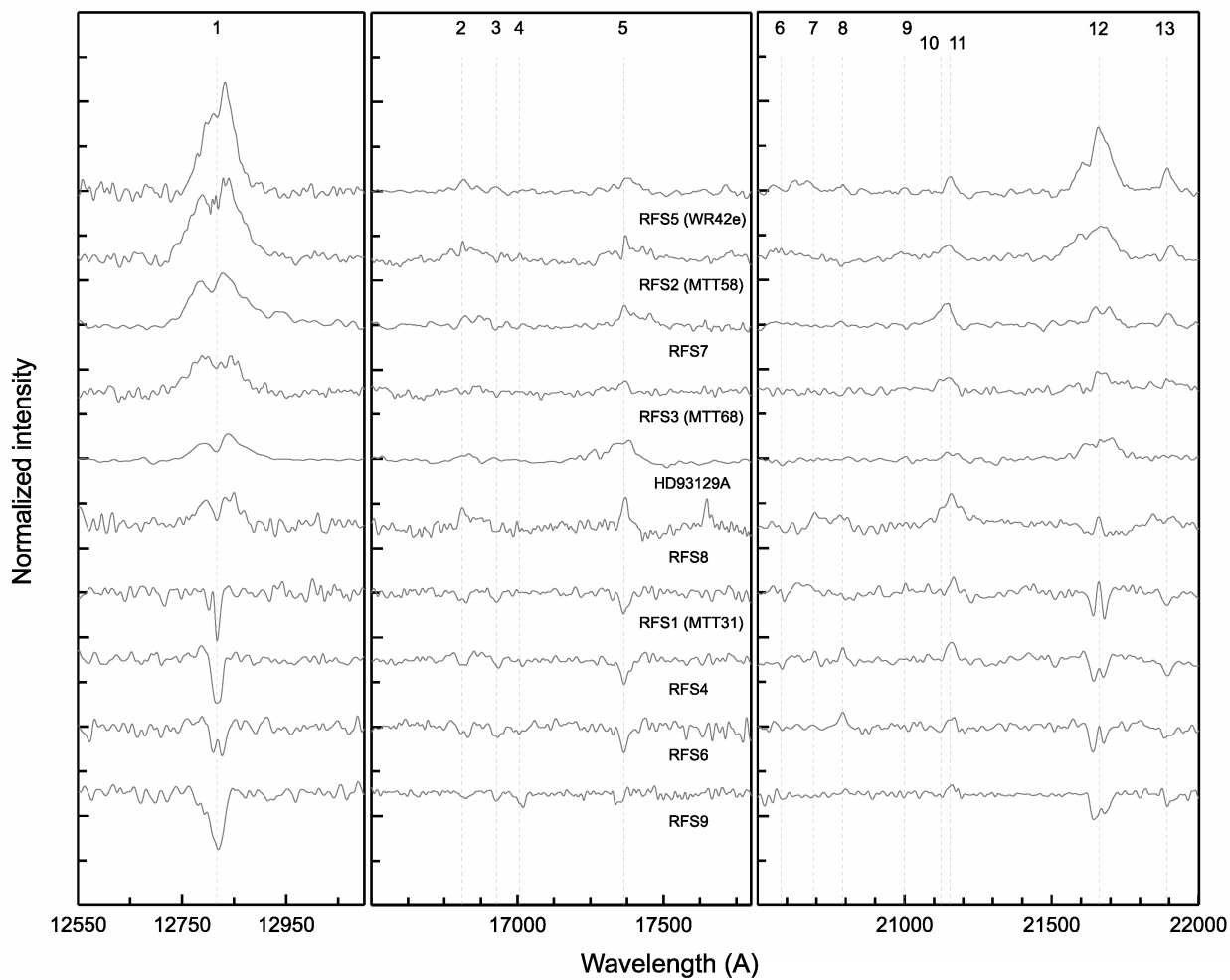}
     \caption{OSIRIS spectra of sources in Table 2. The main line features are indicated as follow: (1) Pa$\beta$ $\lambda$12822, (2) Br 11 $\lambda16811$, (3) He {\sc ii} $\lambda$16930, (4) He {\sc i} $\lambda$17007, (5) Br 10  $\lambda$17367, (6) He {\sc i} $\lambda$20590, (7-8) C {\sc iv} $\lambda$$\lambda$20690-20802, (9) N {\sc v} $\lambda21000$, (10) He {\sc i} $\lambda21126$ (11) N {\sc iii} $\lambda$21160, (12) Br$\gamma$, (13) He {\sc ii} $\lambda$21890.}
        \label{FigVibStab}
  \end{figure}   
%-----------------------------------------------------------------------------------------------------------------------------------------------------------

\subsection{OSIRIS Near-infrared and Goodman Blue-optical spectroscopic dataset}

The NIR spectroscopic data were obtained on observing runs (summarized on Table 3) during nights that in general presented good weather conditions.
The raw frames were reduced following standard NIR reduction procedures, which are presented in details in \citet{roman09}, and shortly described here.
The two-dimensional frames were sky-subtracted for each pair of images taken at two nod positions, followed by division
of the resultant image by a master flat. 
The multiple exposures were combined, and then followed by one-dimensional extraction of the spectra. Thereafter, wavelength calibration was applied 
using the sky lines and being conservative we estimate as $\sim$12-16\AA, the 1-$\sigma$ error for such calibrations. This result comes from the fact 
that we considered the associated 1-$\sigma$ error as the mean values measured from the full width half maximum (FWHM) of the sky lines detected 
in all three NIR bands. On the other hand, if we instead assume the most common criterium of 1/3 of the observed FWHM, then the associated 
1-$\sigma$ error drops to $\sim$4-5\AA.
Also, the effects of the earth atmosphere in the science spectra were corrected using J-, H- and K-band spectra of A-type stars, with the intrinsic
hydrogen lines being carefully subtracted by modeling the observed line profiles through the use of Voigt profiles given by the SPLOT task on IRAF\footnote{http://iraf.noao.edu/}.
The final NIR spectra were normalized through the fitting to the continuum emission observed in the associated wavelength range.
We notice that with the exception of RFS2, all sources in Table 2 were observed in the blue-optical  window in March 2015, with the data being acquired 
using the 1.03$\arcsec$ long slit and the 930 - m2 (3850-5550\AA) grating, which provides a maximum resolving power R$\sim$2100.  On the other hand, RFS2 was 
observed with Goodman in February 2013 using the GG385-600 l/mm grating with the same slit, a setup that provides a maximum resolving power R$\sim$1800, and 
a wavelength coverage 4500-6700\AA.

The journal of the SOAR-Goodman observations is shown in Table 3, and the reduction of the optical spectra was performed using standard techniques through the use of the packages (beside others) ONEDSPEC, TWODSPEC and APEXTRACT within IRAF. 
The one-dimensional spectra of the science targets were extracted from the two dimensional frames by summing pixels in the data range and subtracting off the background value for each column, with the background for each column being measured as the median of non-target pixels for each column. Cosmic rays and other anomalous signal detection were suppressed from each of the extracted spectra, by removing pixels that deviate 5-$\sigma$ of the mean within a 100 pixel wide box that steps through the spectrum. The bad pixels were replaced through a linear interpolation of the removed data range, and the wavelength calibration was performed using Hg(Ar) + Ne lamp spectra. As was made in the case of the NIR spectroscopic observations, the final optical spectra were also 
normalized through the fitting of the continuum emission in the associated wavelength range.

%\textbf{In the next sections we present results obtained from the spectral types, absolute magnitudes and extinction law parameters obtained for the stars in Table 2.}

%-----------------------------------------------------------------------------------------------------------------------------------------------------------
%Figure3
   \begin{figure}
 %   \vspace{0pt}
    %\hspace{25pt}
  \centering
   \includegraphics[width=16.5 cm]{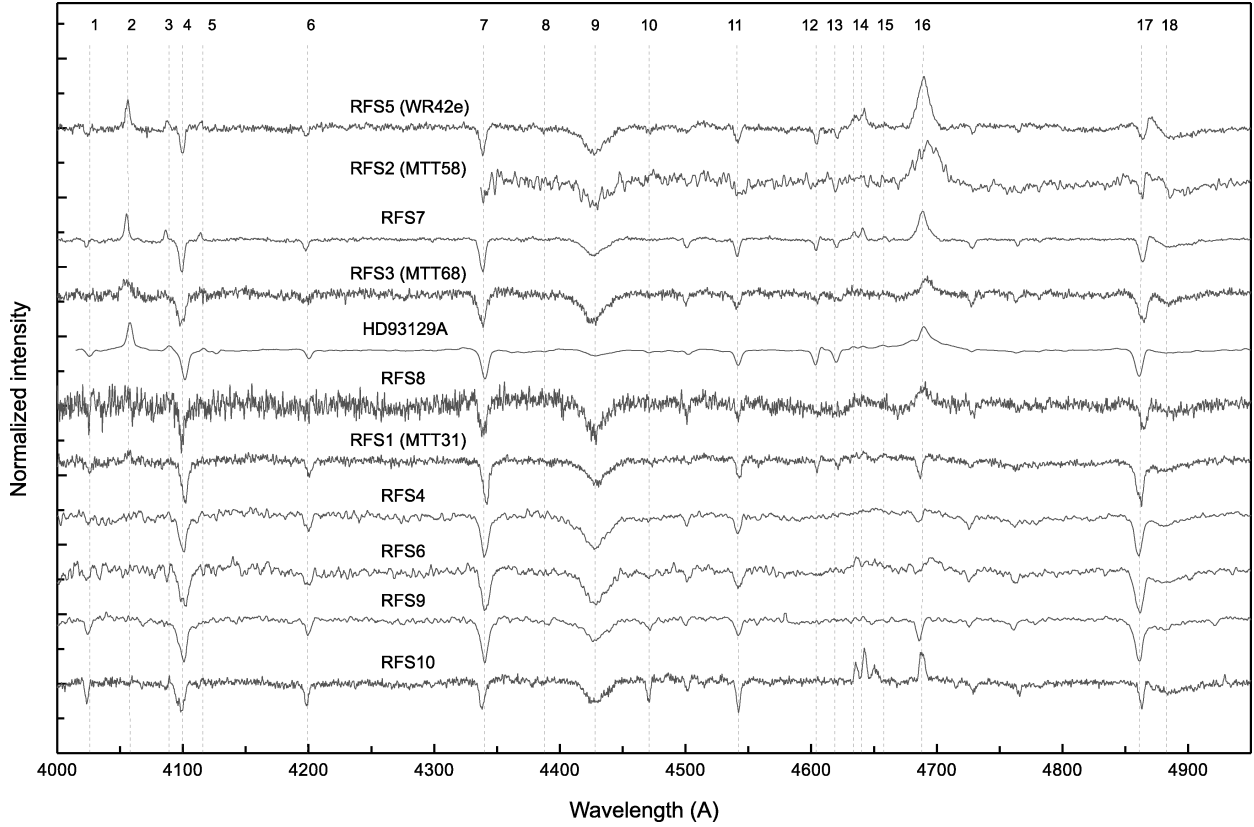}
     \caption{Goodman spectra of sources in Table 2. The main line features are indicated as follow \citep{walb02}: (1) He {\sc i}+{\sc ii} $\lambda$4026, (2) N {\sc v} $\lambda$4058, (3) Si {\sc iv} $\lambda$4089, (4)  H$\delta$, (5) Si {\sc iv} $\lambda$4116, (6) He {\sc ii} $\lambda$4200, (7) H$\gamma$, (8) He {\sc i} $\lambda$4387, (9) DIB $\lambda$4429, (10) He {\sc i} $\lambda$4471, (11) He {\sc ii} $\lambda$4541, (12-13)  N  {\sc v} $\lambda$$\lambda$4604-4620, (14) N  {\sc iii} $\lambda$$\lambda$4634-4642 , (15) C {\sc iv} $\lambda$4658, (16) He {\sc ii} $\lambda$4686, (17) H$\beta$ $\lambda$4861, (18) DIB $\lambda$4882.}
        \label{FigVibStab}
  \end{figure}   
%-----------------------------------------------------------------------------------------------------------------------------------------------------------

%\clearpage

\section{Results and Discussion}

In Figures 2 and 3, we show the OSIRIS and Goodman normalized spectra of the sources in Table 2, with the exception of RFS10 for which we only 
have Goodman optical spectrum. The main NIR and optical spectral features are labeled by numbers, with the corresponding transitions being presented on 
the caption of the figures. As a complement, we also include there the NIR and optical spectra of HD93129A, the prototype of the O2If* class, and until 
recently the unique known Galactic exemplar of this extreme O-type star with blue optical spectra published.

\subsection{Spectral types}

%-----------------------------------------------------------------------------------------------------------------------------------------------------------
%Figure4
   \begin{figure}
 %   \vspace{0pt}
    %\hspace{25pt}
  \centering
   \includegraphics[width=10.5 cm]{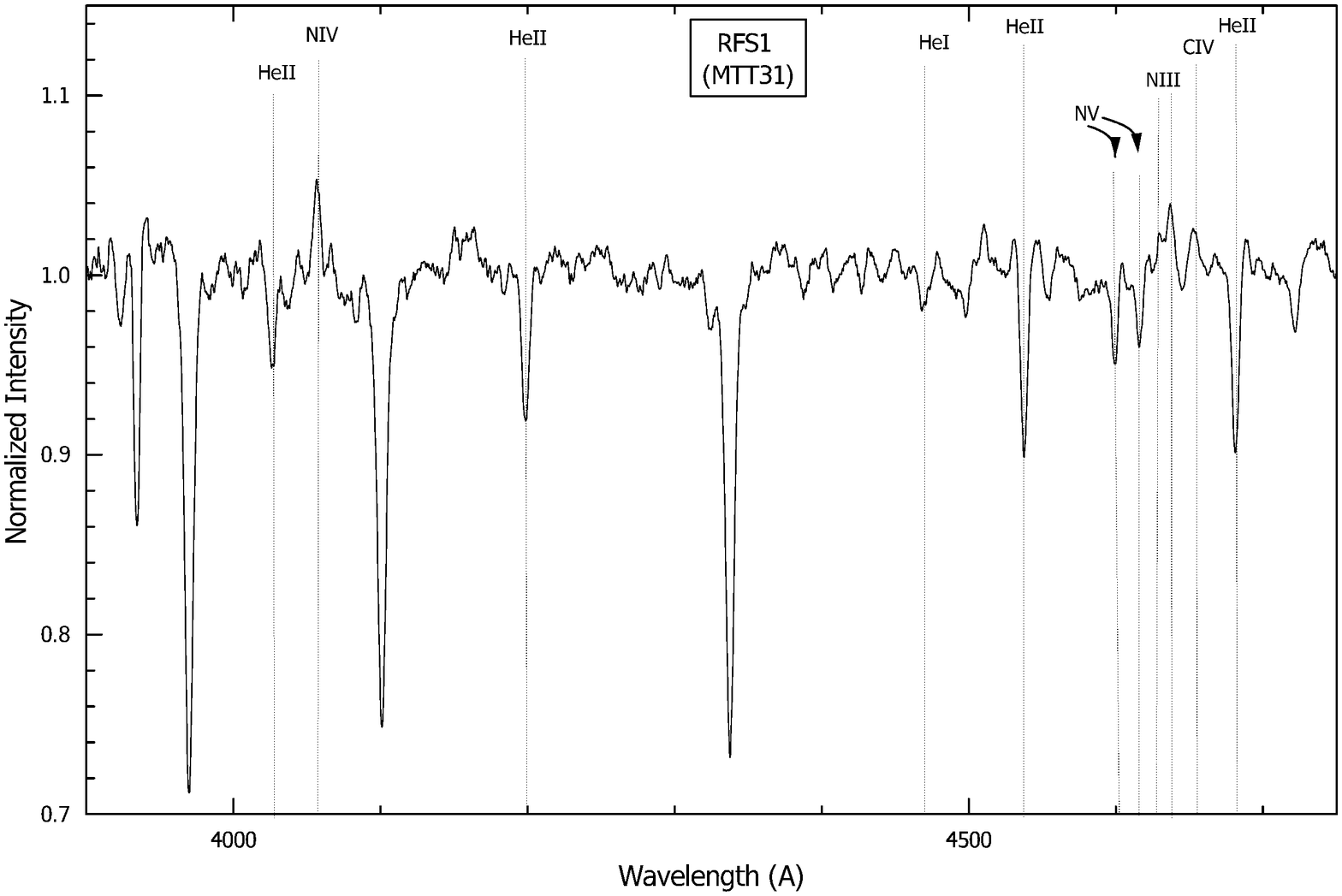}
     \caption{Goodman spectrum of RFS1 source. The main line features are indicated as follow \citep{walb02}: (1) He {\sc i}+{\sc ii} $\lambda$4026, (2) N {\sc v} $\lambda$4058, (3) He {\sc ii} $\lambda$4200, (4) He {\sc i} $\lambda$4471, (5) He {\sc ii} $\lambda$4541, (6)  N  {\sc v} $\lambda$$\lambda$4604-4620, (7) N  {\sc iii} $\lambda$$\lambda$4634-4642 , (8) C {\sc iv} $\lambda$4658, (9) He {\sc ii} $\lambda$4686}
        \label{FigVibStab}
  \end{figure}   

\subsubsection{Three new Galactic exemplars of the OIf*/WN intermediate type confirmed}

The intermediate spectral type OIf*/WN was introduced about three decades ago by \citet{walb82} to classify the emission line star Sk-67 22 in the LMC, which shows spectral features between those of HD93129A (O2If*) and WR20b (WN6ha). 
The OIf*/WN type can be separate from the OIf* and WNH types by the P-Cygni morphology of the optical H$\beta$ line since it is seen uniquely in absorption for O stars (including OIf* stars) and purely in emission for WN stars (Crowther and Walborn 2011). From our NIR survey we identified three new Galactic exemplars of this class,  RFS2 (MTT58), RFS5 (WR42e) and RFS7. As can be seen in Figure 2, their NIR spectrograms are characterized by the presence of strong Pa$\beta$, Br$\gamma$ and He {\sc ii} emission lines. On the other hand, from the optical spectra shown in Figure 3 we notice that they all (but RFS2 whose optical spectrum did not cover this wavelength range) present strong N {\sc iv} $\lambda$4058 lines in emission, with all three also having powerful He {\sc ii} $\lambda$4686 emission lines, as well as H$\beta$ lines showing P-Cygni profiles. Individual comments on the three new OIf*/WN Galactic stars are given as follow:

\paragraph{RFS5} As can be seen in Figure 3, RFS5 has the N {\sc iv} $\lambda$4058 line intensity similar to those of the N {\sc iii} $\lambda$$\lambda$4604-4620 lines, and taking into account the He {\sc ii} $\lambda$4686 line purely in emission an O3If* supergiant type is assigned.
Based on the observed H$\beta$ P-Cygni line profile, and following the criteria presented by \citet{crow11}, we are now able to refine the classification of this source by assigning it the O3If*/WN6 type.  

\paragraph{RFS2} From the RFS2's spectrograms shown in Figures 2 and 3, we can see that they resemble well those of RFS5. The P-Cygni profile in the H$\beta$ line is also evident, confirming the previous classification based on NIR data made by \citet{roman13a}, who classified it as a star of the OIf*/WN (O2If*/WN6) intermediate type. From the new optical data, it is now possible to improve this classification. Indeed, from the comparison of the strength of the N {\sc iii} $\lambda$$\lambda$4604-4620 lines of both stars, which are much weaker in the RFS2's spectrogram than those seen in RFS5, we assign an O2If*/WN5 spectral type to this star.

\paragraph{RFS7} From Figure 2, we can see that RFS7 presents NIR spectral features similar to those seen in the spectrograms of RFS3. Indeed, the intensity (and morphology) of the Pa$\beta$ and Br$\gamma$ lines are quite similar. However, one also can notice that the He {\sc ii} $\lambda$21890 line appears much stronger in the K-band spectrum of RFS7, a characteristic also noticeable in the K-band spectra of RFS2 and RFS5, which combined with the Pa$\beta$ broad emission line morphology, may be useful as complementary criteria when discriminating between the two types using solely NIR spectrograms. In this sense, the P-Cygni profile seen in the RFS7's H$\beta$ line indicates that this star is a new Galactic exemplar of the OIf*/WN type. As a comment on this, the DIB seen at $\lambda$4882 appears relatively weak when compared with those present in the RFS5, RFS2 and RFS3 optical spectrograms. Curiously, the other strong DIB observed at $\lambda$4429 are all more or less of the same intensity, which might indicate that its P-Cygni profile should seem stronger than the one observed. Accordingly to this, the observed line profile would be the result of the combination of two features: a relatively intense H$\beta$ P-Cygni profile suppressed by a strong and broad DIB absorption line at $\lambda$4882. 
Finally, the N {\sc iv} $\lambda$4058 emission line stronger than the N {\sc iii} $\lambda$$\lambda$4604-4620 lines, indicates an intermediate type between O2If*-O3If* \citep{crow11}, so we classify RFS7 as a O2.5If*/WN6 star.      

\subsubsection{RFS1 - The first Galactic exemplar of the O2V class identified to date}

The NIR spectra of RFS1 (MTT31 - presented in Figure 2), are characterized by the presence of weak Pa$\beta$, and Br$\gamma$ hydrogen recombination lines, 
which have peculiar absorption+emission profiles (probably indicative of presence of an intense stellar wind), similar to the one observed in the K-band 
spectrum of Cyg OB2 \#7 \citep{hanson05}. Also, from a careful inspection of the $\lambda$$\lambda$20600-21200 spectral range one may see the existence 
of the C {\sc iv} $\lambda$20802 line in absorption, together with the N{\sc v} $\lambda$21000 line in emission, features normally only seen in K-band 
spectrograms of stars as early as HD64568 (O3 {\sc v}((f))) and Cyg OB2 \#7 (O3If*). Moving to the H-band, we notice that the He {\sc i} $\lambda$17007 
line is absent in the RFS1's spectrum. However, it does show a strong He {\sc ii} $\lambda$16930 line in absorption, as well as the He {\sc ii} $\lambda$21890 
in the K-band, indicating that RFS1 is a very hot star probably earlier than O3.\\
In Figure 3 it is shown the Goodman blue-optical spectrum of RFS1, and in order to allow a better analysis of it, in Figure 4 it is presented its individual 
spectrum in a more detailed and useful way. In particular, the weak lines close to the He {\sc ii} $\lambda$4686 line are better seen in this version of the 
spectrum. The N {\sc iv} $\lambda$4058 line is seen in emission, and the strong He {\sc ii} absorption line at 4200\AA, confirms the impression from our above 
discussion, e.g. that RFS1 is a very hot O-type star. In fact, the N {\sc iv} $\lambda$4058, and N {\sc iii} $\lambda$$\lambda$4634-4642 emission lines give 
strong support to this idea. Finally, the C {\sc iv} $\lambda$4658 line clearly detected in emission led us to classify RFS1 as a new O2{\sc v} 
star \citep{walb02}. Indeed, based on a careful inspection of the blue-optical spectrum of BI 253 (the prototype of the O2{\sc v} class \citep{walb02}) 
shown in figure 12 (a) of \citet{massey05}, we can see that the RFS1 blue-optical spectrum is a nearly clone of it, which makes this star the first know 
Galactic exemplar of the O2{\sc v} type \citep{sota14}.

On the other hand, the detection of what probably is a very weak He {\sc i} $\lambda$4471 absorption line may suggests the presence of a later O-type 
companion. 
From a search in the literature, we found that RFS1 has an associated X-ray point source at only 0.16$\arcsec$ from its coordinates, catalogued as 
CXOU111506.69-611633.4 \citep{townsley14}. However, the observed X-ray to bolometric luminosity ratio ($\sim$ 3$\times$10$^{-8}$) is found to be 
compatible with what is expected in case of a single star (L$_X$/L$_{\odot}$ $\sim$10$^{-7}$). Further photometric and spectroscopic monitoring 
studies are necessary in order to test this assumption.

\subsubsection{RFS3 confirmed as a new Galactic O2If* star}

RFS3 (MTT68) was first catalogued as a probable member of NGC 3603 by \citet{melnick89}, and discovered to be a strong Chandra point source 
by \citet{moffat02}, who performed an extensive X-ray to radio study of the 3.6$\arcmin$$\times$3.6$\arcmin$ field centered on NGC 3603. Later, 
based on J-, H- and K-band SOAR-OSIRIS spectra it was classified as a O2If* star by \citet{roman13a}, who found its NIR spectra very similar to 
those of HD93129A, the template of the class \citep{walb02} and at that time, the only Galactic O2If* previously known.
Based on our new Goodman blue-optical spectra of RFS3 (shown in Figure 3), we may now confirm that this object is indeed a highly reddened O2If* star, 
as indicate the strong diffuse interstellar bands (DIBs) seen at $\sim$ 4430\AA, and 4480\AA. As previously noticed by \citet{roman13a} in the NIR, the 
RFS3 optical spectrum also looks pretty similar to that of HD93129A, as indicate the intense N {\sc iv} $\lambda$4058 emission line much stronger than 
the N {\sc iii} emission line pair at $\lambda$$\lambda$4634-4642, which combined with the strong He {\sc ii} $\lambda$4686 emission line \citep{crow11}, 
confirms its O2If* nature.
On the other hand, from a careful inspection of the NIR emission line features seen in the spectra of both stars (e.g. the Pa$\beta$ and the 
N {\sc iii} $\lambda$21150 lines), we can see that the observed emission lines appear broader in the RFS3 NIR spectrograms, a characteristic that is 
also quite evident in the N {\sc iv} $\lambda$4058 optical line, which may suggest the existence of a secondary companion of RFS3, which in this case 
should also be another extreme early-type star.

\subsubsection{New Early- and Mid-O Galactic supergiants}

Besides the O2If* and OIf*/WN stars discussed above, there are two other sources in Table 2 (RFS8 and RFS10) that show the He {\sc ii} $\lambda$4686 line in emission. Unfortunately, due to a problem during the acquisition process, the NIR spectra of RFS10 resulted not useful for spectral analyses, so in its case our main conclusions only rely on optical data. In what follows, we discuss in some detail the results obtained for this two objects.

\paragraph{RFS8} The NIR spectrograms of the RFS8 source (shown in Figure 2), are dominated by emission line features associated to the Pa$\beta$, Br$\gamma$, Br10, and Br11 hydrogen atomic transitions, as well as by broad emission line profiles corresponding to the C {\sc iv} $\lambda$$\lambda$20690-20802 and N {\sc iii} $\lambda$21160 transitions. The morphology of the Pa$\beta$ emission line is quite similar (on both shape and intensity) to the one seen in the J-band spectrum of HD93129A, however, the Br$\gamma$ line profile is completely different. Also, the He {\sc ii} lines $\lambda$16930 and $\lambda$21890 appear in absorption, suggesting a later spectral type. 
The blue-optical spectrum for this source (shown in Figure 3) confirms this assumption. The intense emission feature associate to He {\sc ii} $\lambda$4686 indicates a supergiant status for this object. The N {\sc iv} $\lambda$4058 emission line is less intense than those of the N {\sc iii} $\lambda$$\lambda$4634-4642 pair, which combined with the N {\sc v} $\lambda$$\lambda$4604-4620 lines weaker than those seen in the HD93129A blue-optical spectrogram, indicate a spectral type later than O3. Finally, the absence of a He {\sc i} $\lambda$4471 absorption line, suggests a spectral type probably no later than O4-O4.5 \citep{sota11}. Taking into account the points discussed above, and the criteria presented by \citet{crow11}, we assign an O3.5If* spectral type to the RFS8 source.

\paragraph{RFS10}
As mentioned before, for this star we only have optical data, and its spectrogram is shown in Figure 3. It is a high S/N optical spectrum in which the He {\sc ii} $\lambda$4686 emission line denotes a supergiant class for this source. The absence of N {\sc v} $\lambda$$\lambda$4604-4620 lines associated to the presence of a strong He {\sc i} $\lambda$4471 absorption line, indicate that this star is of spectral type later than O3.5-O4 \citep{sota11}. Also, from the library of optical spectra of O-stars of \citet{sota11}, one can see that the relative intensity of the two N {\sc iii} $\lambda$$\lambda$4634-4642 emission lines and the He {\sc i} $\lambda$4471 and He {\sc ii} $\lambda$4541 absorption lines, make this spectrum pretty similar to that of HD 169582 the standard of the O6Ia type.
We searched in the literature looking for X-ray sources associated to RFS10 founding a strong one in the XMM-Newton catalogue at only 0.35$\arcsec$ from RFS10's coordinates, identified as J111955.1-611603. Further photometric and spectroscopic studies are indicated for this source, as the detected X-ray emission could be due to the presence of a close binary companion.

\subsubsection{early-O type stars}

The sources in this group are RFS4, RFS6 and RFS9, whose spectral features are typical of early and mid-O type stars. In the following, we discuss each object in detail.

\paragraph{RFS4 (MTT71)} 

The J-, H-, K-band spectra of RFS4 (presented in Figure 2) are very similar to those of RFS1, with a small but remarkable difference: the presence 
of He{\sc i} $\lambda$17007 absorption line, which however is less intense than the He{\sc ii} $\lambda$16930 line. This feature indicates that this 
source is also an early O-type star, and from a comparison with the H-band spectra of HD46223 (O4{\sc v}), HD66811 (O4 I), HD14947 (O5 If) and 
Cyg OB2 8c (O5 If) \citep{hanson05}, one may see that the H- and K-band features of RFS4 are compatible with those seen in O-type stars of spectral types O4-O5. 
On the other hand, the Goodman normalized spectrum of RFS4 (shown in Figure 3), does not show the He{\sc i} $\lambda$4471 absorption line, so an early O-type 
status for this star is confirmed. Also, the absence of a N{\sc iv} $\lambda$4058 line (in emission or absorption) indicates that this star is of spectral type 
not earlier than O4. Indeed, the optical spectrum of HD168076AB (O4{\sc iii}) is virtually a clone of the RFS4 Goodman spectrogram, including the incipient 
P-Cygni profile seen in its He{\sc ii} $\lambda$4686 line, allowing us to classify RFS4 as a new Galactic O4{\sc iii} star. 
As a final comment on this source, from a cross check with X-Ray catalogues in the literature, we found a point source (at only 0.05$\arcsec$ from it) 
catalogued as CXOU111521.31-611504.3 \citep{townsley14}. Indeed, RFS4 was first discovered as a strong X-ray point source by \citet{moffat02} who pointed that 
it could be due to colliding wind binary interaction, which may suggest the existence of an early O-type companion. Further studies are necessary in order to 
properly address this assumption. 

\paragraph{RFS6}

The OSIRIS and Goodman spectra for this source look pretty similar to those of the RFS4 source. Indeed, the presence of strong He{\sc ii} $\lambda$$\lambda$16930-21890 absorption lines, the weakness of He{\sc i} $\lambda$17007 transition combined with the absence of the He{\sc i} absorption line $\lambda$21130
give support to this idea. Also, taking into account the presence of strong C{\sc iv} and N{\sc iii} $\lambda$$\lambda$20800-21160 emission lines led us to the conclusion that this star is probably another exemplar of the early type O-star family. In support to that, the optical spectrogram shown in Figure 3 also looks pretty similar to that of RFS4, except for the He{\sc i} $\lambda$4471 line which is clearly seen (weakly) in absorption, which is not uncommon for O4-5 stars like HD46223 (O4{\sc v}), HD66811 (O4{\sc i}), HD14947 (O5{\sc i}f) and Cyg OB2 8c (O5{\sc i}f) \citep{hanson05}. 
%Finally, as was the case for RFS4, there is also a X-Ray point source,  CXOU111453.56-612423.0 \citep{townsley14} at only 0.3$\arcsec$ from the RFS6 coordinates, which may also indicate a binary nature.

\paragraph{RFS9}

The NIR spectrograms of the RFS9 source are shown in Figure 2. As can be seen there, its J-band spectrum presents a strong and broad Pa$\beta$ absorption line, a feature also noticed on the profiles of the H- and K-band Bracket lines. The He {\sc ii} lines at $\lambda$16930 and $\lambda$21890 are also seen in absorption, and there is no sign of the He  {\sc i} line at 21113\AA, so a spectral type later than O6{\sc v} is ruled out \citep{hanson05}. On the other hand, the He {\sc i} absorption line $\lambda$17007 is stronger than the He {\sc ii} $\lambda$16930, an indication that this star is possibly of a type later than O4. From the optical spectrogram of the RFS9 source, shown in Figure 3, we can see that the He {\sc ii} $\lambda$4686 line is seen purely in absorption, so a dwarf class is assigned. Based on the above discussion, we performed a careful examination of O4{\sc v}-O5{\sc v} optical spectra presented in the literature, concluding that our Goodman spectrum of the source RFS9 looks pretty similar to those of HD15629 (O4.5{\sc v}), HD46150 (O5{\sc v}) and HD93204 (O5{\sc v}), presented in \citet{sota11}, and that of source \#108 (O5.5{\sc v}) of \citet{melena08}.
An interesting spectral characteristic that kept our attention is the He {\sc ii} $\lambda$$\lambda$4686 absorption line stronger than any other He {\sc ii} lines present in the $\lambda$$\lambda$4000-5000 wavelength range. 
Such O-type spectral feature is believed to be characteristic of the Vz class. It seems that the Of effect (established as a luminosity indicator in normal O-type spectra) is suppressed by the Vz effect, generating less emission than in normal class V spectra, which might indicate extreme stellar youth \citep{walb14}.

%-----------------------------------------------------------------------------------------------------------------------------------------------------------
%Figure 5
   \begin{figure}
 %   \vspace{0pt}
    %\hspace{25pt}
  \centering
   \includegraphics[width=8.5 cm]{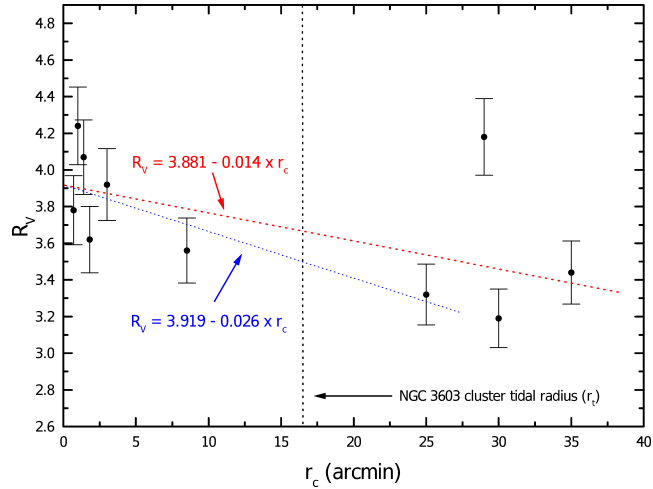}
     \caption{R$_V$ as a function of the cluster center radius r$_c$ for sources in Table 4. We can see that with the exception of RFS8, the associated R$_V$ values decreases approximately linearly as a function of r$_c$. In the same plot we also present the results of linear fittings considering all sources in the sample, as well as those (six) presenting r$_c$ values less than the NGC 3603 cluster tidal radius estimated by \citet{sung04}.}
        \label{FigVibStab}
  \end{figure}   
%-----------------------------------------------------------------------------------------------------------------------------------------------------------

%-----------------------------------------------------------------------------------------------------------------------------------------------------------
%Figure 6
   \begin{figure}
 %   \vspace{0pt}
    %\hspace{25pt}
  \centering
   \includegraphics[width=8.5 cm]{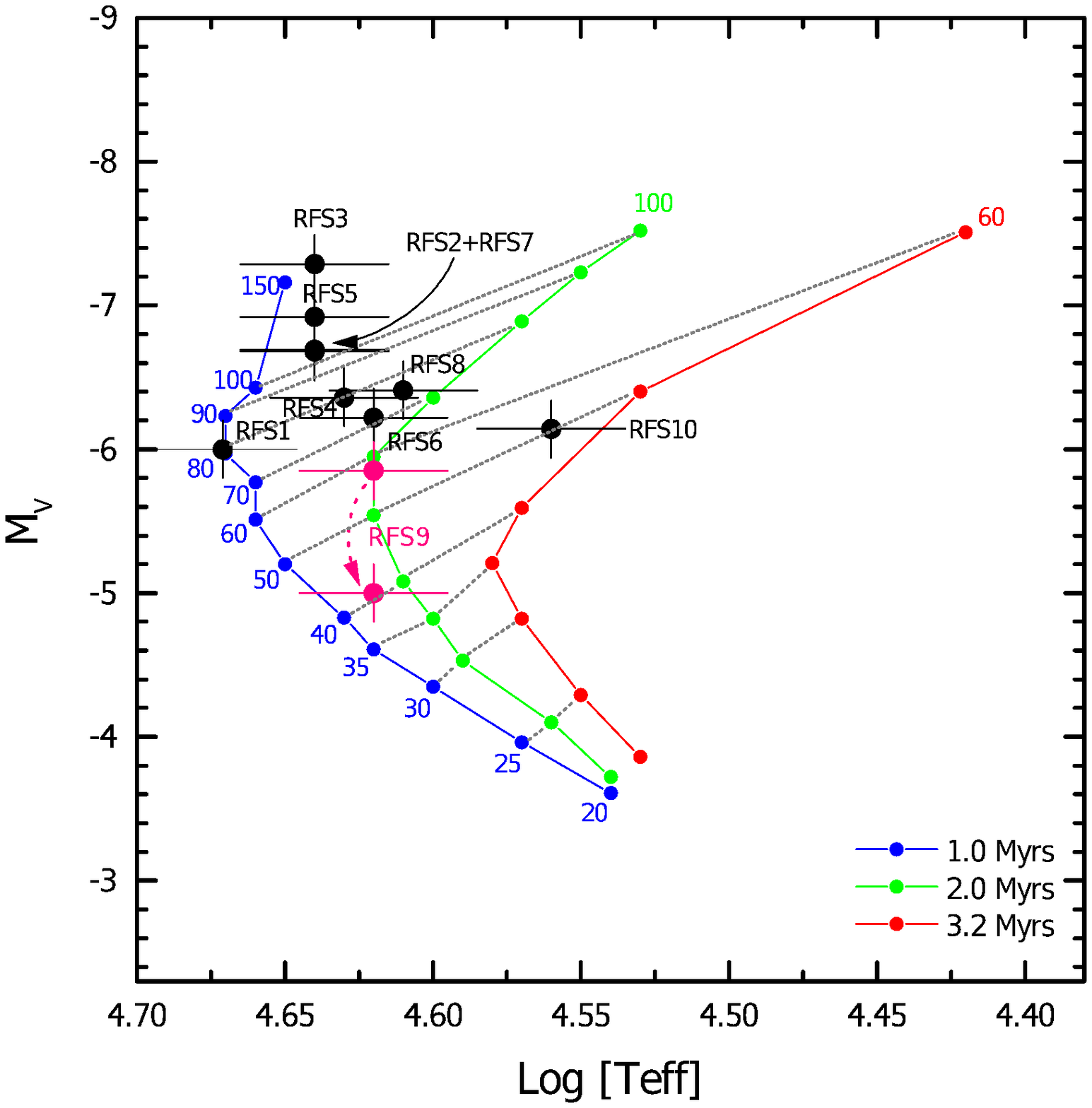}
     \caption{The M$_V$ versus Log T (Teff) Hertzprung-Russel diagram for sources in Table 4. The effective temperatures for the OI, OIII and OV classes 
     were computed using Equation 2 presented in the work of \citet{martins05}, while the values for the OIf*/WN stars were estimated from the work 
     of \citet{crow10}. For all sources we assumed an uncertainty of $\pm$3000K on the quoted temperature values. We also present there the 1 Myr, 
     2 Myrs and 3.2 Myrs isochrons taken from the work of \citet{bressan12}, for stars with initial 
     masses in the range 20-150M$_{\odot}$(solar metallicity Z=0.015). RFS9 is represented considering single and binary systems comprised by two 
     stars of same spectral type. In this sense, assuming a mean absolute visual magnitude M$_V$ = -5.0 for each component would result in a combined 
     absolute magnitude M$_V$ = -5.8, a value in line with the one derived by us (considering the associated uncertainties) using the distance modulus 
     equation, equivalent to a binary system comprised by two stars of 40 M$_{\odot}$ each (see text in Section 4.2.2).}
        \label{FigVibStab}
  \end{figure}   
  %\footnote{http://stev.oapd.inaf.it/cgi-bin/cmd}
%-----------------------------------------------------------------------------------------------------------------------------------------------------------

A natural question that arises from this result is, how such an extremely young O-star could be found at such a large angular distance from the NGC3603 cluster center? A possible answer is that RFS9 belongs to an independent star forming complex embedded in the main body of the NGC3603 complex. In an upcoming paper (Roman-Lopes et al. in preparation), we further discuss the issue on how so many of the sources in Table 2 could have arrived at their present location.

%\clearpage
%-----------------------------------------------------------------------------------------------------------------------------------------------------------
%Figure 7
   \begin{figure}
 %   \vspace{0pt}
    %\hspace{25pt}
  \centering
   \includegraphics[width=13.0 cm]{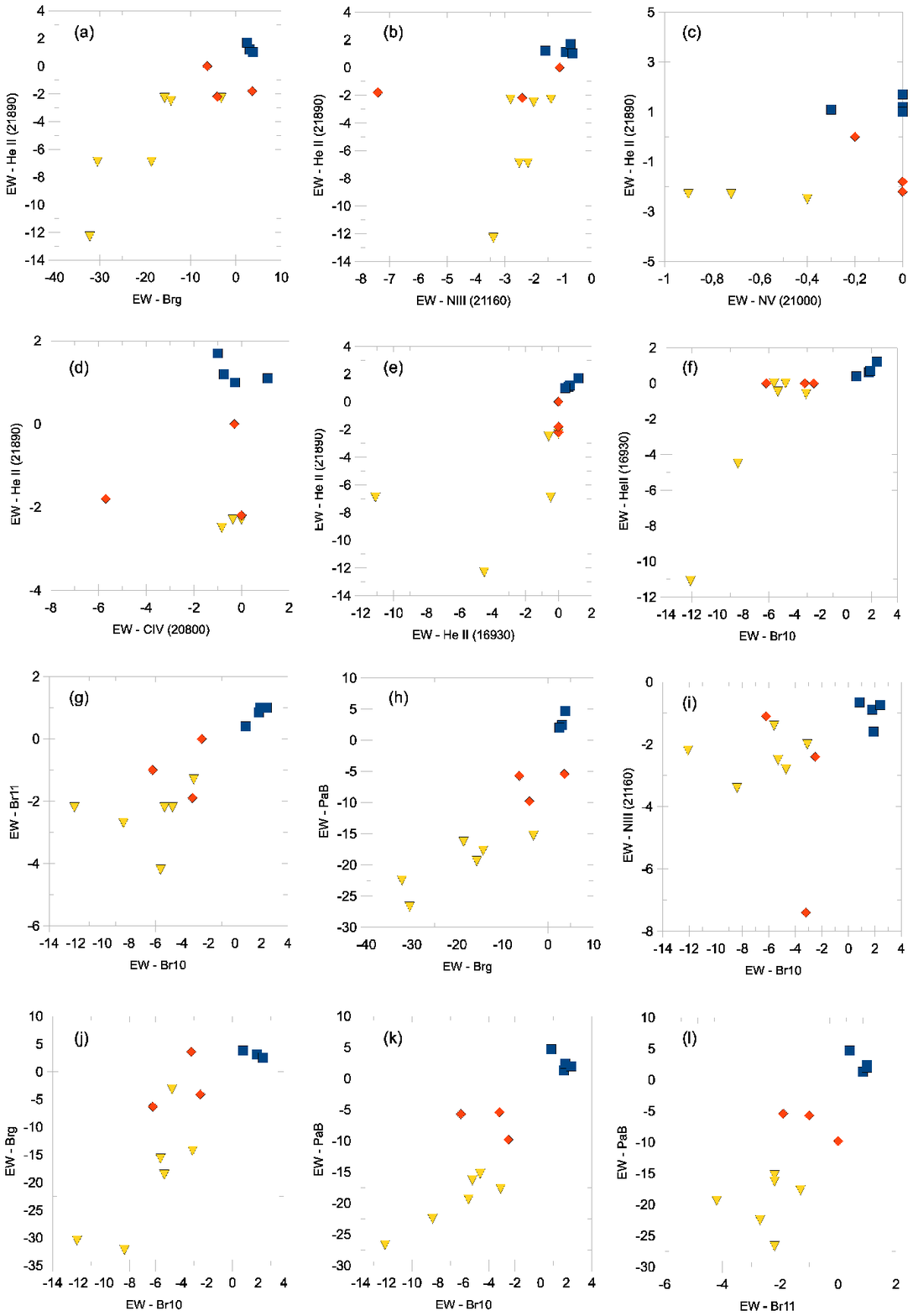}
     \caption{Observed values of EWs (\AA) for O (squares), OIf* (diamonds) and OIf*/WN type  (triangles) stars in Table 5. As can be noticed from panels 7(e) to 7(l), the use of H- and J-bands spectral line measurements can also provide a good separation of the distinct types. In fact, the combination of the Br10, Br11 and Pa$\beta$ hydrogen line measurements is particularly useful when separating the transitional OIf*/WN stars from the OIf* ones. In this sense, good results are achieved using Pa$\beta$ line measurements combined with the H-band hydrogen transitions.}
        \label{FigVibStab}
  \end{figure}
%-----------------------------------------------------------------------------------------------------------------------------------------------------------

%\clearpage
%-----------------------------------------------------------------------------------------------------------------------------------------------------------
%Figure 8
   \begin{figure}
 %   \vspace{0pt}
    %\hspace{25pt}
  \centering
   \includegraphics[width=14.5 cm]{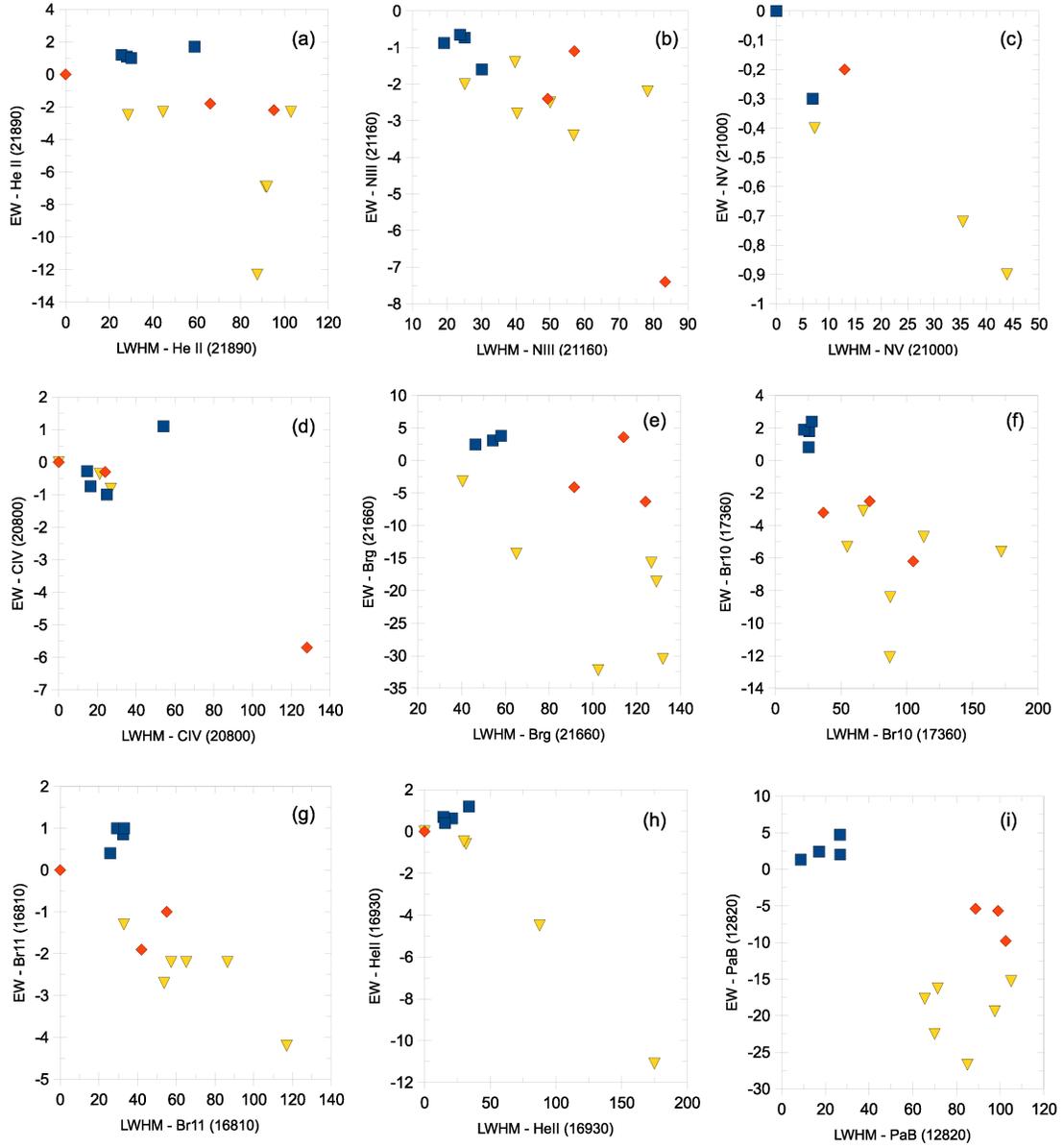}
     \caption{Observed values of EW $\times$ LWHM (\AA) for O (squares), OIf* (diamonds) and OIf*/WN type  (triangles) stars in Table 5.}
        \label{FigVibStab}
  \end{figure}   
%-----------------------------------------------------------------------------------------------------------------------------------------------------------

%\clearpage
%-----------------------------------------------------------------------------------------------------------------------------------------------------------
%Figure 9
   \begin{figure}
 %   \vspace{0pt}
    %\hspace{25pt}
  \centering
   \includegraphics[width=8.5 cm]{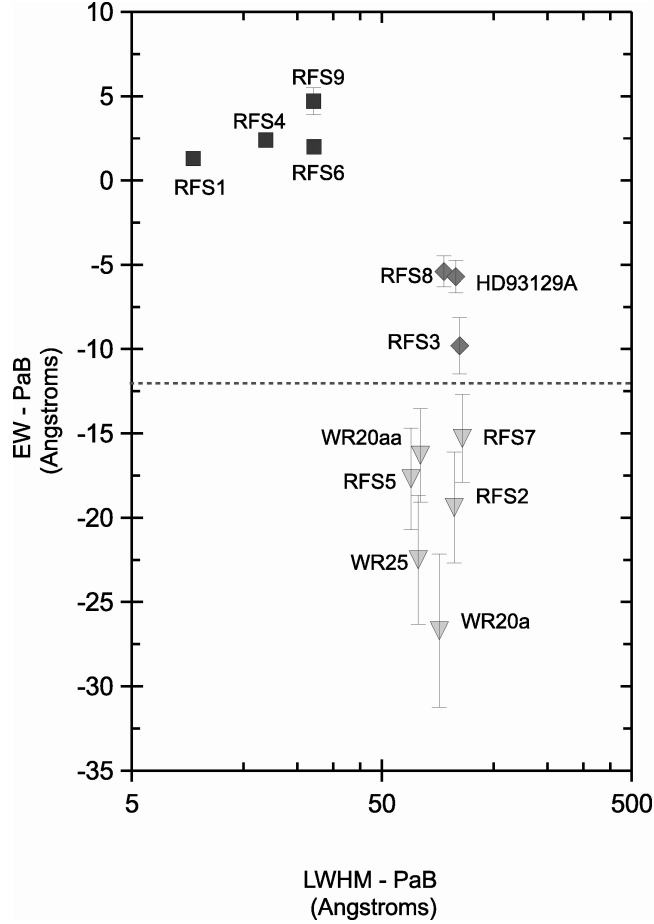}
     \caption{EW $\times$ LWHM Pa$\beta$ values for O (squares), OIf* (diamonds) and OIf*/WN type  (triangles) stars in Table 5. The results for the Pa$\beta$ emission line indicate that the transition boundary between OIf* and OIf*/WN stars occurs for EW (Pa$\beta$) $\sim$ 12\AA, always for spectral lines showing LWHM (Pa$\beta$) values above $\sim$ 60\AA.}
        \label{FigVibStab}
  \end{figure}

%-----------------------------------------------------------------------------------------------------------------------------------------------------------

\subsection{Interstellar extinction, absolute magnitudes, ages and masses} 

%-----------------------------------------------------------------------------------------------------------------------------------------------------------

%%Table 4 

\begin{deluxetable}{ccccccccccccccccccc}
\tabletypesize{\scriptsize}
\rotate
\tablecaption{Spectral types, absolute magnitudes and extinction law parameters obtained for sources in Table 2. (1) ID, (2-6) (B-V) colors and associated optical to NIR color excesses. The color-excesses were calculated using the corresponding (B-V)$_0$, (J-V)$_0$,  (H-V)$_0$ and (K-V)$_0$ intrinsic colors taken from \citet{martins06}.  (7) The total to selective extinction ratio R$_V$ values were derived using the equations A3 to A6 of \citet{fitz99}. (8) Visual extinction computed from E(B-V) and R$_V$. (9) Absolute visual magnitudes M$_V$ obtained using A$_V$ plus the visual magnitudes for each source in Table 2, applied into the distance modulus equation for an heliocentric distance of 7.6$\pm$0.35 kpc \citep{crow10}. (10-12) The same for the M$_J$, M$_H$ and M$_K$ absolute magnitudes, using the V-J, V-H and V-K color values taken from \citet{martins06}. (13-15) A$_J$, A$_H$ and A$_K$ extinction values derived using the J-, H- and K-band magnitudes from Table 2, together with the associated M$_J$, M$_H$ and M$_K$ values applied into the distance modulus equation.}
\tablewidth{0pt}
\tablehead{\colhead{Source} & SpType & Log Teff & \colhead{(B-V)}& \colhead{E(B-V)} & \colhead{E(J-V)} & \colhead{E(H-V)} & 
\colhead{E(K-V)} & \colhead{R$_V$} & \colhead {A$_V$}  & \colhead{M$_V$} &  \colhead{M$_J$} & \colhead{M$_H$} &\colhead{M$_K$} & \colhead{A$_J$} &\colhead{A$_H$} & \colhead{A$_K$}  \
%  & & EW -- LW & EW -- LW & EW -- LW & EW -- LW & EW -- LW & EW -- LW & EW -- LW  & EW -- LW  & EW -- LW  & EW -- LW & EW -- LW 
}
\startdata

RFS1 & O2 V & 4.67 & 1.22 & 1.50 & -4.07 & -4.69 & -5.09 & 3.78 & 5.67 & -6.00 & -5.33 & -5.21 & -5.12 & 1.60 & 0.98 & 0.58  \\
RFS2 & O2 If*/WN5 & 4.64 & 1.38 & 1.66 & -4.96 & -5.87 & -6.40 & 4.24 & 7.04 & -6.68 & -6.01 & -5.89 & -5.80 & 2.08 & 1.17 & 0.64  \\
RFS3 & O2 If* & 4.64 & 1.59 & 1.87 & -5.41 & -6.34 & -6.86 & 4.07 & 7.61 & -7.29 & -6.62 & -6.50 & -6.41 & 2.20 & 1.27 & 0.75  \\
RFS4 & O4 III & 4.63 & 1.57 & 1.85 & -4.81 & -5.54 & -6.02 & 3.62 & 6.70 & -6.36 & -5.69 & -5.57 & -5.48 & 1.89 & 1.16 & 0.68  \\
RFS5 & O3 If*/WN6 & 4.64 & 1.52 & 1.80 & -5.02 & -5.85 & -6.37 & 3.92 & 7.05 & -6.92 & -6.25 & -6.13 & -6.04 & 2.03 & 1.20 & 0.68  \\
RFS6 & O4.5 V & 4.62 & 1.50 & 1.78 & -4.55 & -5.25 & -5.67 & 3.56 & 6.33 & -6.22 & -5.55 & -5.43 & -5.34 & 1.78 & 1.08 & 0.66  \\
RFS7 & O2.5 If*/WN6 & 4.64 & 1.28 & 1.56 & -3.71 & -4.29 & -4.65 & 3.32 & 5.18 & -6.69 & -6.02 & -5.90 & -5.81 & 1.47 & 0.89 & 0.53  \\
RFS8 & O 3.5If* & 4.61 & 1.32 & 1.60 & -4.74 & -5.55 & -6.08 & 4.18 & 6.69 & -6.41 & -5.74 & -5.62 & -5.63 & 1.95 & 1.14 & 0.61  \\
RFS9 & O5 Vz & 4.61 & 1.36 & 1.64 & -3.74 & -4.35 & -4.67 & 3.19 & 5.22 & -5.85 & -5.18 & -5.06 & -4.97 & 1.48 & 0.87 & 0.55  \\
RFS10 & O6 Ia & 4.56 & 1.35 &1.63  & -3.99 & -4.62 & -5.08 & 3.44 & 5.60 & -6.14 & -5.47 & -5.35 & -5.26 & 1.61 & 0.98 & 0.52  \\

\enddata
%\tablecomments{SpT corresponds to the spectral types of the sources, as derived in previous work using NIR data}.  
%A portion is 
%shown here for guidance regarding its form and content.}
%\tablenotetext{a}{Sample footnote for table~\ref{tbl-1} that was generated
%with the deluxetable environment}
%\tablenotetext{a}{The column SpType presents the spectral types derived in previous studies using only NIR spectra}
\end{deluxetable} 

\subsubsection{Interstellar extinction}

The interstellar extinction law in the direction of NGC 3603 appears to be anomalous. This is not surprisingly as it is well known that very young 
H{\sc ii} regions in general show ratio of total to selective extinction values well above the canonical value R$_V$=3.1 \citep{chini83,chini90}.
\citet{pandey00} performed an extensive UBVRI CCD photometric survey in the direction of the innermost region (2.08' $\times$ 3.33' FOV) of the cluster, and 
from a variable extinction analysis of the sources with color excess in the range 1.30 $<$ E(B-V) $<$ 1.75, they computed a mean value for the total to 
selective extinction ratio R$_V$=3.8$\pm$0.8. The large uncertainty on the mean reflects the large scatter on the individual derived values. Also, by 
following the method of \citet{neckel81} they were able to obtain the intra-cluster total to selective extinction ratio value R$_V$=4.1$\pm$0.2.

In Table 4 besides the spectral types and effective temperatures, we present the (B-V) colors for each star in our sample, together with the 
associated optical to NIR color excesses E(B-V), E(J-V),  E(H-V) and E(K-V), which were derived using the intrinsic colors for the earliest O-stars 
given by \citet{martins06}. The values of the total to selective extinction ratio were then estimated using the equations A3-A6 of \citet{fitz99}, 
and by using R$_V$ and E(B-V), we computed the corresponding extinction parameter A$_V$ for each star in Table 2. From the R$_V$ values in Table 4, 
we can see that for most of the stars in our sample the associated interstellar law is in agreement with the previous study of \citet{pandey00}, 
with R$_V$ ranging from 3.19 for RFS9 to 4.24 in case of RFS2, with mean value and standard deviation of the mean R$_V$=3.73$\pm$0.35. As in the 
previous studies, the relatively large error value on the mean probably reflects the moderate scatter on the computed ratio of total to selective 
extinction in the direction of NGC 3603, which in turn is probably produced by the variations on the dust and gas column densities, and perhaps on 
differences on dust composition and grain sizes. 

In Figure 5, we plot the parameter R$_V$ as a function of the cluster center distance r$_c$ for sources in Table 4. We can see that with the exception 
of RFS8, the associated R$_V$ values decreases as a function of r$_c$. In the same plot we also present the results of linear fittings considering all sources 
in the sample, as well as only those (six) presenting r$_c$ values less than the NGC 3603 cluster tidal radius estimated by \citet{sung04}. The resulting linear 
equations correlating R$_V$ and r$_c$ are R$_V$ = 3.881 - 0.014r$_c$ and R$_V$ = 3.919 - 0.026r$_c$, for the entire sample and only for those sources inside the 
NGC 3603 tidal radius area, respectively.
The values of R$_V$ computed from our fittings limited to the radial cluster center distance of 2 arcmin are R$_V$ = 3.85 and R$_V$ = 3.87, respectively, in good 
agreement with the result obtained by \citet{pandey00}. Finally, from the extinction parameters presented in Table 4, we computed mean values and uncertainties 
(errors on the mean) of the ratios A$_\lambda$/A$_V$ for the J-, H- and K$_S$-bands obtaining A$_J$/A$_V$ = 0.286$\pm$0.004, A$_H$/A$_V$ = 0.170$\pm$0.003 and 
A$_{Ks}$/A$_V$ = 0.098$\pm$0.005.

\subsubsection{Absolute magnitudes, ages and masses}

The absolute magnitudes for sources in Table 4 were derived using the observed magnitudes and associated extinction values applied into the distance 
modulus equation, adopting an heliocentric distance of 7.6$\pm$0.35 kpc \citep{crow10}. Also, the respective M$_J$, M$_H$ and M$_K$ absolute magnitudes were 
calculated using the (V-J)$_0$, (V-H)$_0$ and (V-K)$_0$ intrinsic colors for early-O stars taken from \citet{martins06}.
In Figure 6, we present the M$_V$ versus Log T (Teff) Hertzprung-Russel (H-R) diagram for sources in Table 4. The effective temperatures for the O{\sc i}, 
O{\sc iii} and O{\sc v} classes were obtained using Equation 2 presented in the work of \citet{martins05}, while the values for the OIf*/WN stars were estimated 
from the work of \citet{crow10}. For all sources we assumed an uncertainty of $\pm$3000K on the quoted values. We also present there the 1 Myr, 2 Myrs and 
3.2 Myrs isochrons\footnote{http://stev.oapd.inaf.it/cgi-bin/cmd} taken from the work of \citet{bressan12}, for stars with initial masses in the range 
20-150M$_{\odot}$(solar metallicity Z=0.015).

From this diagram we can see that the youngest sources are RFS1, RFS2, RFS3, RFS5 and RFS7 with probable ages of about 1 Myrs. The new O2If* star (RFS3) 
with initial mass above 150 M$_{\odot}$, appears as the most massive and luminous object of our sample. This is not surprising as O2If* stars are known to belong 
to the most massive and luminous type of O-stars in the local universe. For example, HD93129A the most massive and luminous O-star previously known in the Galaxy 
has an absolute visual magnitude M$_V$=-6.6 \citep{simon83} and estimated initial mass of 130$\pm$15 M$_{\odot}$ \citep{taresch97}. On the other hand, in the LMC 
we have the extreme case of Mk42, an O2If* star presenting absolute visual magnitude M$_V$=-7.4, and initial mass of 189 M$_{\odot}$ \citep{besten14}.

Accordingly with the isochrons shown in Figure 6, all three new Galactic OIf*/WN stars had initial masses well above 100 M$_{\odot}$, with 
RFS5 (O3If*/WN6) being a bit more massive than RFS2 and RFS7. Placed in the same star forming complex, NGC3603-C has the same spectral type (O3If*/WN6), 
with M$_V$=-7.2 and estimate initial mass in the range 123-154 M$_{\odot}$ \citep{crow10}. Taking into account the evolutionary models and the RFS5's position 
in the H-R diagram of Figure 5, we can conclude that for an estimated age of 1 Myr, RFS5 should have initial and current masses of 135 M$_{\odot}$ and 
123 M$_{\odot}$, respectively, a result in line with those obtained by \citet{crow10} in their analyses of NGC3603-C. Besides the three new OIf*/WN stars, 
another very young star in our sample is RFS1. From its position in the H-R diagram of Figure 6, we can see that it is an O2{\sc v} star with probable initial 
and current masses of 80 M$_{\odot}$ and 76 M$_{\odot}$, respectively. As the only Galactic star of its class known to date, RFS1 has an absolute magnitude 
M$_V$=-6.0, corresponding to a luminosity similar to other O2{\sc v} stars found in the LMC, like VFTS 621 (O2{\sc v} and VFTS 512 (O2{\sc v}-{\sc iii}) 
both presenting visual absolute magnitudes M$_V$=-6.1 \citep{besten14}.

A second group of sources slightly older with ages ranging from 1.5-2.0 Myrs, is comprised by RFS4 (O4{\sc iii}), RFS6 (O4.5{\sc iii}), 
RFS8 (O3.5If*) and RFS9 (O5Vz). From their associated absolute visual magnitudes (Table 4) we can see that they appear also as very luminous sources, 
and based on the comparison of such values with those from other stars of similar spectral types, we conclude that our results are in line with those 
found in the literature, with the exception of RFS9, which appears more luminous than what is expected from the comparison with other LMC and Galactic 
O5{\sc v}z stars. We will discuss more about this source later. 

The absolute visual magnitudes of RFS4 and RFS6 are quite consistent with the values for stars of similar spectral types found in the LMC, as 
for example VFTS 422 (O4{\sc iii}) - M$_V$=-5.8, VFTS 603 (O4{\sc iii}) - M$_V$=-6.3, and VFTS 608 (O4{\sc iii}) - M$_V$=-6.0. On the other hand, 
in the Milk Way we have the cases of HD 168076 (O4{\sc iii}) with absolute visual magnitude M$_V$=-5.9 \citep{sana09}, HD 93250 AB (O4{\sc iii}) 
presenting M$_V$=-6.2 with an estimated stellar mass of 83 M$_{\odot}$\citep{repol04}, as well as in the innermost part of NGC 3603, the 
source \#38 of \citet{melena08} (O4{\sc iii}) with M$_V$ = -6.0.

The case of RFS8 (O3.5If*) is quite interesting. Indeed, it is located well to the south of the complex, in apparent isolation at the large radial 
angular center distance of 29', or equivalently, at a projected linear distance of about 62 pc considering the quoted NGC 3603 heliocentric distance of 7.6 kpc. 
Based on its location in the H-R diagram of Figure 5 and accordingly with the stellar evolutionary models, RFS8 would have initial and current masses of 
77 M$_{\odot}$ and 70 M$_{\odot}$, respectively. Its derived absolute magnitude M$_V$=-6.4 is similar to that of another O3.5If* (Sh18) found in the NGC 3603's 
innermost region, with an absolute magnitude M$_V$=-6.3 \citep{besten11}. The fact that a such high mass star is found isolated in the field naturally led us to 
speculate if it could have been expelled sometime in the past from the innermost parts of the complex. Looking for proper motion measurements for this star and 
based on a search on the PPMXL and XPM proper motion catalogues \citep{roeser10,fedo11}, we found that RFS8 has proper motion parameters 
pmRA=-34.3$\pm$11.8 mas/yr  - pmDE=-53.2$\pm$11.8 mas/yr and  pmRA=-40.0$\pm$10.0 mas/yr  - pmDE=-58.6$\pm$10.0 mas/yr, respectively, quite high for an object 
that is assumed to be placed at the same heliocentric distance of NGC 3603. Indeed, considering a distance of 7.6$\pm$0.35 kpc would results in a projected 
linear velocity of about 240 km/s! 
Despite such a high velocity for a very high mass star at first might seen unlikely, \citet{gvara11} showed from their study on the dynamical ejection scenario of 
massive runaway stars that such a high mass runaway star could actually exist. Indeed, they argue that the most effective process generating this kind of high 
velocity star is that produced by a close fly-by dynamical encounter between a single massive star with
a very massive hard binary system. For example, in the case of a NGC3603-A1-like binary system (M$_1$ = 120  M$_\odot$ + M$_2$ = 90 M$_\odot$), in a few percent 
of the cases a 60-80 M$_\odot$ M$_3$ fly-by star would acquire a peculiar velocity as high as 250 km/s. In this sense, further proper motion measurements like 
those to be provided by the Gaia space astrometry mission, are necessary to give (or not) additional support to this idea.

From RFS9's location in the HR diagram of Figure 5, we can see that its derived absolute magnitude (M$_V$ = -5.85) is not consistent with the absolute visual 
magnitudes for single stars of the O5{\sc v}z type in the LMC, like for example VFTS 385 (M$_V$ = -5.2, log (Teff) = 4.63), VFTS 511 (M$_V$ = -4.9, log (Teff) = 4.64), 
and VFTS 581 (M$_V$ = -5.0, log (Teff) = 4.60) \citep{sabin14}. A possible explanation for this discrepancy can be obtained if we consider the case in which RFS9 is a 
binary system comprised by two stars of same spectral type. In this sense, assuming a mean absolute visual magnitude M$_V$ = -5.0 for each component would result in a 
combined absolute magnitude M$_V$ = -5.8, a value in line with the one derived by us (considering the associated uncertainties) using the distance modulus equation. 
In this case, from Figure 5 we can see that RFS9 would be a binary system comprised by two stars of 40 M$_\odot$ each.

With an estimated age of about 3 Myrs (accordingly with the isochrons of \citet{bressan12}), RFS10 appears as the oldest star in our sample with initial 
and current masses of 45-55 M$_\odot$ and 40-50 M$_\odot$, respectively. Its derived absolute visual magnitude M$_V$ = -6.14 and spectroscopic mass of 
45-55 M$_\odot$ agree with the corresponding values obtained for Galactic O5-O6{\sc i}a stars (M$_V$ = -6.1) \citep{wegner05}, as well as with the values of 
stars of similar spectral type and luminosity class in the LMC, like VFTS 151 (O6.5{\sc ii}(f)p - M$_V$ = -6.4, M = 79 M$_\odot$), 
VFTS 208 (O6 (n)fp - M$_V$ = -5.8, M = 53 M$_\odot$) and VFTS 440 (O6-O6.5{\sc ii}(f) - M$_V$ = -6.2, M = 76 M$_\odot$).

\subsection{Identifying the OIf*/WN type solely from near-IR spectrograms}

With the usage of new efficient detectors and large ground-based telescopes, the near-IR window has become available for spectral typing (\citet{hanson96,hanson05,crow11}, and references therein). This is particularly important when dealing with highly obscured and distant early-type stars. However, despite the observational progress made during the last years, the number of known Galactic OIf*/WN stars with some study in both, optical and near-IR domains is relatively small. Considering that the majority (if not all) of the remaining members of the
OIf*/WN Galactic stellar population probably is going to be discovered (and properly studied), mainly from the use of NIR spectroscopic facilities, 
one important question (already made by \citet{crow11}) is, can we safely distinguish between the OIf* and OIf/WN types using only near-IR spectroscopy?
In order to contribute to address this question, in the next we will make use of the NIR spectral line measurements taken from our new set of Galactic O, OIf* and OIf*/WN stars, which were spectroscopically classified using the data taken with OSIRIS and Goodman at SOAR.

In Tables 5 and 6 we present the equivalent width (EW) and line width (LW) measurements made from the observed NIR and optical lines. As a complement 
we also present in Table 5 the EW and LW measurements of relevant NIR lines, obtained from a set of un-published J-, H- and K-band OSIRIS spectra of WR20a 
(O3If*WN6+O3If*/WN6), WR20aa (O2If*/WN5) and WR25 (O2.5If*/WN6+O) \citep{ross15}, taken during previous nights with SOAR.
From the associated values, we constructed EW and EW$\times$LWHM comparative diagrams shown in Figure 7 and Figure 8, respectively, and based on them, one 
can see that the normal O-type stars (represented by the blue squares) in general are seen well separated from the OIf* and OIf*/WN stars. This is particularly 
true even when considering spectral lines seen in emission in all NIR spectrograms, like for example the N {\sc iii} $\lambda$$\lambda$21160 transition (panel 7(b)). 
On the other hand, the separation of the normal O-type stars from the OIf* and OIf*/WN groups is clearly seen in all remaining panels of Figure 7. 
 
In the particular case of using only the K-band to separate normal O-type stars from the OIf* and OIf*/WN types, the best comparative diagram to be used is the one 
presenting the He {\sc ii} $\times$ Br$\gamma$ EW line measurements, as seen in panel 7(a). In fact, in this case it is possible not only to identify the locus of 
normal O-type stars, but also to separate the majority of the OIf*/WN type stars from the OIf* group. This is particularly evident for those OIf*/WN stars 
presenting He {\sc ii} $\lambda$$\lambda$21890 and Br$\gamma$ line width values satisfying simultaneously the condition EW(He {\sc ii}) $<$ -2\AA, and 
EW(Br$\gamma$) $<$ -12.5\AA, which correspond to a combined width W$_\lambda$(Br$\gamma$+He {\sc ii}) $\sim$ 15\AA, about half of the value suggested 
by \citet{crow11} for the criterium defining the boundary between OIf* and OIf*/WN stars.
On the other hand, as can be noticed from panels 7(e) to 7(l), the use of H- and J-bands spectral line measurements can also provide a good 
separation of the distinct types. In fact, the combination of the Br10, Br11 and Pa$\beta$ hydrogen line measurements is particularly useful when separating 
the transitional OIf*/WN stars from the OIf* ones. In this sense, good results are achieved using Pa$\beta$ line measurements combined with the H-band hydrogen 
transitions, as can be seen in panels 7(k) for EW (Pa$\beta$) $\times$ EW (Br10) and 7(l) for EW (Pa$\beta$) $\times$ EW (Br11). 

From such diagrams one may conclude that the approximate values defining the boundaries between OIf* and OIf*/WN stars are those corresponding to combined equivalent widths EW$_\lambda$(Br$\gamma$+pa$\beta$) $>$ 18\AA (panel 7(h)),  EW$_\lambda$(Br10+pa$\beta$) $>$ 20\AA (panel 7(k)), and EW$_\lambda$(Br11+pa$\beta$) $>$ 16\AA (panel 7(l)).
 Finally, regarding approximate boundaries between subtypes analogous
to the criterium using the He II $\lambda$4686 in the optical window \citep{crow11}, the results for the Pa$\beta$ emission line are shown in Figure 9. There it is clearly seen that the transition boundary between OIf* and OIf*/WN stars occurs for EW (Pa$\beta$) $\sim$ 12\AA, always for lines showing LWHM (Pa$\beta$) values above $\sim$ 60\AA. 
 
%%Table 5

\begin{deluxetable}{cccccccccccc}
\tabletypesize{\scriptsize}
\rotate
\tablecaption{Equivalent width (EW) and NIR line width (LW) measurements (\AA), for sources in Table 2, with the exception of RFS10 for which we only have optical 
data, with the uncertainty on the quoted values varying from 10\% to 15\%.}
\tablewidth{0pt}
\tablehead{\colhead{Source} & \colhead{Pa$\beta$ (1)}& \colhead{Br11 (2)} & \colhead{He {\sc ii} (3)} & \colhead{He {\sc i} (4)} & 
\colhead{Br10 (5)} & \colhead{C {\sc iv} (7-8)} & \colhead{N {\sc v} (9)} & \colhead{He {\sc i} (10)} & \colhead{N {\sc iii} (11)} & 
\colhead {Br$\gamma$ (12)} & \colhead{He {\sc ii}} (13) \\
 & EW -- LW & EW -- LW & EW -- LW & EW -- LW & EW -- LW & EW -- LW & EW -- LW  & EW -- LW  & EW -- LW  & EW -- LW & EW -- LW \
} 
\startdata
%\data
RFS1 & 1.3 --  8.8 & 0.85 --  32.5 & 0.62 --  21 & 0 --  0 & 1.8 --  25.9 & 1.1 --  54 & -0.3 --  6.9 & 0 --  0 & -0.88 --  19.2 &  em+abs   & 1.1 --  27.9 \\
RFS2 & -19.4 --  97.5 & -4.2 --  117 & 0 --  0 & 0 --  0 & -5.6 --  172 & 0.36 --  21.2 & -0.72 --  35.5 & 0 --  0 & -1.4 --  39.8 & -15.7 --  126.7 & -2.3 --  44.5 \\
RFS3 & -9.8 --  102.5 & 0 --  0 & 0 --  0 & 0 --  0 & -2.5 --  71.9 & 0 --  0 & 0 --  0 & 0 --  0 & -2.4 --  49.3 & -4.1 --  91.4 & -2.2 --  95.2 \\
RFS4 & 2.4 --  17.2 & 1 --  29.4 & 0.7 --  14.4 & 0.2 --  8.5 & 1.9 --  21.9 & -0.74 --  16.5 & 0 --  0 & 0 --  0 & -1.6 --  30.2 & 3.1 --  54.2 & 1.2 --  25.5 \\
RFS5 & -17.7 --  65.5 & -1.3 --  32.9 & -0.59 --  31.6 & 0 --  0 & -3.1 --  67 & -0.82 --  26.9 & -0.4 --  7.3 & 0 --  0 & -2 --  25.2 & -14.3 --  65 & -2.5 --  28.5 \\
RFS6 & 2.0 --  26.8 & 1 --  33.0 & 1.2 --  33.9 & 0.9 --  40.4 & 2.4 --  27.8 & -1 --  24.8 & 0 --  0 & 0 --  0 & -0.73 --  25.2 & 2.5 --  46.2 & 1.7 --  58.9 \\
RFS7 & -15.3 -- 105 & -2.2 --  86.4 & 0 --  0 & 0 --  0 & -4.7 --  113 & 0 --  0 & -0.9 --  43.9 & 0 --  0 & -2.8 --  40.4 & -3.2 --  40.4 & -2.3 --  103 \\
RFS8 & -5.4 --  88.7 & -1.9 --  42 & 0 --  0 & 0 --  0 & -3.2 --  36.6 & -5.7 --  128 & 0 --  0 & 0 --  0 & -7.4 --  83.4 & 3.6 --  114 & -1.8 --  66.1 \\
RFS9 & 4.7 --  26.7 & 0.4 --  25.9 & 0.4 --  15.5 & 1.1 --  22.1 & 0.82 --  25.5 & -0.28 --  14.7 & 0 --  0 & 0 --  0 & -0.65 --  23.8 & 3.8 --  58 & 1 --  30 \\
HD93129A & -5.7 --  99.0 & -1 --  55 & 0 --  0 & 0 --  0 & -6.2 --  105 & -0.3 --  24 & -0.2 --  13 & 0 --  0 & -1.1 --  57 & -6.3 --  124 & 0 --  0 \\
WR20a & -26.7 -- 85.0 & -2.2 -- 65.1 & -11.1 -- 175 &    & -12.1 -- 87.2 &    &    &    & -2.2 -- 78.3 & -30.5  -- 132 & -6.9 -- 91.4 \\
WR20aa & -16.3 -- 71.4 & -2.2 -- 57.3 & -0.47 -- 30 &    & -5.3 -- 54.8 &    &    &    & -2.5 -- 50 & -18.6 --129 & -6.9 -- 92 \\
WR25 & -22. 5 -- 70.0 & -2.7 -- 53.7 & -4.5 -- 87.6 &    & -8.4 -- 87.6 &    &    &    & -3.4 -- 56.8 & -32.2 --102.4 & -12.3 -- 87.6 \\

\enddata
%\tablecomments{SpT corresponds to the spectral types of the sources, as derived in previous work using NIR data}.  
%A portion is 
%shown here for guidance regarding its form and content.}
%\tablenotetext{a}{Sample footnote for table~\ref{tbl-1} that was generated
%with the deluxetable environment}
%\tablenotetext{a}{The column SpType presents the spectral types derived in previous studies using only NIR spectra}
\end{deluxetable}


%% Included in this acknowledgments section are examples of the
%% AASTeX hypertext markup commands. Use \url without the optional [HREF]
%% argument when you want to print the url directly in the text. Otherwise,
%% use either \url or \anchor, with the HREF as the first argument and the
%% text to be printed in the second.

%%Table 6

\begin{deluxetable}{ccccccccccccc}
%\begin{table}{ccccccccccccc}
\tabletypesize{\scriptsize}
\rotate
\tablecaption{Equivalent width (EW) and line width (LW) measurements of selected optical lines (\AA), for sources in Table 2, with the uncertainty on the quoted values ranging from 10\% to 15\%.  We also show the values for HD 93129A taken from the optical spectrum in \citet{roman11} }
\tablewidth{0pt}
\tablehead{
\colhead{Source} & \colhead{He {\sc i}+{\sc ii} (1)}& \colhead{N {\sc iv} (2)} & \colhead{He {\sc ii} (6)} & \colhead{He {\sc i} (10)} & 
\colhead{He {\sc ii} (11)} & \colhead{N {\sc v} (12)} & \colhead{N {\sc v} (13)} & \colhead{N {\sc iii} (14-1)} & \colhead{N {\sc iii} (14-2)} & 
\colhead {C {\sc iv} (15)} & \colhead{He {\sc ii}} (16) & \colhead{H$\beta$} (17) \\
 & EW -- LW & EW -- LW & EW -- LW & EW -- LW & EW -- LW & EW -- LW & EW -- LW  & EW -- LW  & EW -- LW  & EW -- LW & EW -- LW & Profile
}
\startdata
%\data
RFS1  &  0.34  --    3  &  -0.46  --    4.5  &  0.57  --    4.4  &  0.06  --    1.02  &  0.71  --    4.1  &  0.23  --    2.1  &  0.18  --    2.5  &  -0.16  --    3.6  &  -0.23  --    3.8  &  -0.16  --    4.7  &  0.71  --    4.1  &  pure abs  \\
RFS2  &    --      &    --      &    --      &    --      &  1.5  --    10.4  &  0.27  --    5.1  &  0.25  --    2.9  &    --      &    --      &    --      &  -7.5  --    20.6  &  P-Cygni  \\
RFS3  &  0  --    0  &  -1  --    7.7  &  0.39  --    6.7  &  0  --    0  &  0.73  --    5.4  &  0.4  --    6.2  &  0.47  --    8.1  &  -0.2  --    14.3  &  0  --    0  &  0  --    0  &  -1.46  --    10.7  &  pure abs  \\
RFS4  &  0  --    0  &  0  --    0  &  0.76  --    5.4  &  0  --    0  &  0.82  --    4.4  &  0  --    0  &  0  --    0  &  0  --    0  &  0  --    0  &  0  --    0  &  0.53  --    5.3  &  pure abs  \\
RFS5  &  0.3  --    3.3  &  -1.2  --    3.6  &  0.33  --    3.6  &  0.14  --    3.3  &  0.52  --    4.4  &  0.38  --    3.2  &  0.26  --    3.2  &  -0.58  --    5.8  &  -0.72  --    4.8  &  0  --    0  &  -6.6  --    16.5  &  P-Cygni  \\
RFS6  &  0  --    0  &  0  --    0  &  0.69  --    5.3  &  0.32  --    6  &  0.99  --    6.4  &    --      &    --      &    --      &    --      &    --      &  0.53  --    5.4  &  pure abs  \\
RFS7  &  0.12  --    2.1  &  -0.8  --    2.8  &  0.34  --    3.8  &  0  --    0  &  0.56  --    3.7  &  0.2  --    2.3  &  0.17  --    2.6  &  -0.19  --    2.8  &  -0.36  --    3.3  &  -0.28  --    4.8  &  -4.3  --    12.8  &  P-Cygni  \\
RFS8  &  0  --    0  &  -0.26  --    1.5  &  0.54  --    6  &  0  --   0  &  0.4  --    2.8  &  0.21  --    6  &  0.07  --    1.9  &  -0.4  --    6  &  -1.1  --    10  &  -0.7  --    11.2  &  -2.2  --    11.3  &  pure abs  \\
RFS9  &  0.58  --    4.1  &  0  --    0  &  0.6  --    3.8  &  0.57  --    5.9  &  0.73  --    5  &  0  --    0  &  0  --    0  &  0  --    0  &  0  --    0  &  0  --    0  &  0.86  --    3.8  &  pure abs  \\
RFS10  &  0.47  --    2.6  &  0  --    0  &  0.8  --    3.9  &  0.59  --    3.2  &  0.98  --    3.3  &  0  --    0  &  0  --    0  &  -0.41  --    2.9  &  -0.88  --    3.4  &  0  --    0  &  -1.28  --    4.4  &  pure abs  \\
HD93129A  &  0.28  --    4.7  &  -1.5  --    5.5  &  0.35  --    4.6  &  0.13  --    5.8  &  0.8  --    4.7  &  0.7  --    4  &  0.67  --    4.7  &  -0.1  --    4.2  &  -0.09  --    4.5  &  -0.09  --    5.4  &  -4.2  --    24  &  pure abs  \\

\enddata
%\tablecomments{SpT corresponds to the spectral types of the sources, as derived in previous work using NIR data}.  
%A portion is 
%shown here for guidance regarding its form and content.}
%\tablenotetext{a}{Sample footnote for table~\ref{tbl-1} that was generated
%with the deluxetable environment}
%\tablenotetext{a}{The column SpType presents the spectral types derived in previous studies using only NIR spectra}
\end{deluxetable}
%\end{table}

\section{Concluding remarks}

In this work we performed a search for very massive star candidates beyond the center of the massive stellar cluster NGC3603, which is known to be one of the most massive, dense and rich 
Galactic star-forming region, and believed to be a scaled version of the R136 star-burst cluster in the LMC.
Based on NIR colour and magnitude selection criteria applied to the 2MASS point source catalogue, the chosen stars were observed through a SOAR NIR and optical spectroscopic survey which 
confirmed the existence of several massive stars in isolation in the NGC3603 field.

From the analysis of the spectroscopic survey and related optical-NIR photometry, our main results are:

%\begin{enumerate}
\begin{itemize}
\item Three new Galactic exemplars of the OIf*/WN type, RFS2 (MTT58), RFS5 (WR42e) and RFS7, are confirmed in the periphery of NGC 3603. They are very young massive stars with probable initial 
masses well above 100 M$_{\odot}$ and estimated ages of about 1 Myr. RFS5 (O3If*/WN6 - 135 M$_{\odot}$) appears to be a bit more massive than RFS2 and RFS7, both with estimated initial masses 
of $\sim$ 115 M$_{\odot}$. Such results are in line with the estimated mass of NGC3603-C, the O3If*/WN6 star placed in the core of the NGC 3603, which has an estimated mass of 123-154 
M$_{\odot}$ and absolute visual magnitude M$_V$=-7.2 \citep{crow10}.

\item Based on a Goodman blue-optical spectrum of RFS3 (MTT68), we may now confirm that it is indeed a highly reddened O2If* star, the only other Galactic exemplar (besides HD93129A) 
known to date. Based on its position relative to a set of theoretical isochrons in a Hertzprung-Russel diagram, we concluded that the new O2If* star is probably one of the most massive 
(150 M$_{\odot}$) and luminous (M$_V$=-7.3) O-star in the Galaxy.

\item The RFS1 blue-optical spectrum is a nearly clone of the prototype of the O2{\sc v} class, the BI 253 star in the LMC  \citep{walb02}), which makes RFS1 the first Galactic exemplar 
known to date. Based on its location on the Hertzprung-Russel diagram, we found that it is a star with probable initial mass of 80 M$_{\odot}$, and luminosity similar to the other O2{\sc v} 
stars found in the LMC.

\item The case of RFS8, a new Galactic O3.5 If* star is quite intriguing. It is found well to the south of the NGC 3603 complex, in apparent isolation at a large radial angular center 
distance of 29', or equivalently, at a projected linear distance of about 62 pc. Based on its location in the H-R diagram and accordingly with the stellar evolutionary models, RFS8 
probably had an initial mass of 77 M$_{\odot}$. Its derived absolute magnitude M$_V$=-6.4 is similar to that of the other O3.5If* (Sh18) found in the NGC 3603's innermost region. 
The fact that a such high mass star is found well isolated in the field led us to speculate that it could have been expelled from the innermost parts of the complex.

\item From the spectroscopic study and associated optical and NIR photometry, we were able to derive the values of the total to selective extinction ratio R$_V$ and the corresponding 
optical to NIR color excesses E(B-V), E(J-V),  E(H-V) and E(K-V) for each star in our sample. Our results are consistent with an anomalous interstellar extinction law found on previous 
studies in the direction of the NGC 3603 region. In this sense, we found a radial dependence on the total to selective extinction ratio, with the associated R$_V$ values decreasing as a 
function of r$_c$ given by R$_V$ = 3.881 - 0.014r$_c$.

\item  Based on the equivalent width (EW) and line width (LW) measurements made from the observed NIR spectroscopic lines, we constructed EW and EW$\times$LWHM comparative diagrams and 
found that the normal O-type stars can be easily separated from the OIf* and OIf*/WN type stars. Also, in the particular case of using only the K-band, it is possible not only to separate 
normal O-type stars from the OIf* and OIf*/WN types, but also to identify the majority of the OIf*/WN type stars that present He {\sc ii} $\lambda$$\lambda$21890 and Br$\gamma$ line width 
values satisfying simultaneously the condition EW(He {\sc ii}) $<$ -2\AA, and EW(Br$\gamma$) $<$ -12.5\AA, which correspond to a combined 
width W$_\lambda$(Br$\gamma$+He {\sc ii}) $\sim$ 15\AA, about half of the value suggested by \citet{crow11}. 

On the other hand, in the absence of K-band spectroscopic data, excellent results in the separation process are achieved using Pa$\beta$ line measurements combined with those of H-band 
hydrogen Br10 and Br11 transitions. In this case the values defining the boundaries between OIf* and OIf*/WN stars correspond to the combined equivalent widths 
satisfying EW$_\lambda$(Br10+pa$\beta$) $>$ 20\AA, and EW$_\lambda$(Br11+pa$\beta$) $>$ 16\AA.

Finally, regarding approximate boundaries between subtypes analogous to the criterium using the He II $\lambda$4686 in the optical window \citep{crow11}, the best choice in the 
NIR is that using the Pa$\beta$ emission line for which we clearly see a transition boundary between OIf* and OIf*/WN stars occurring for EW (Pa$\beta$) $\sim$ 12\AA and 
LWHM (Pa$\beta$) values above $\sim$ 60\AA.

%\end{enumerate}
\end{itemize}

\acknowledgments

ARL thanks partial support from DIULS Regular project PR15143. GAPF is partially supported by CNPq and FAPEMIG. We thanks the anonymous referee by the careful
reading of the manuscript. Her/his criticism was appreciated.
We thanks the SOAR staff for the efficient support provided during the OSIRIS and Goodman observing runs. ARL thanks Dr. Nolan Walborn by stimulating discussions about optical spectra of O2V stars.
ARL thanks Dr. Nidia Morrel by stimulating discussions on optical spectral classification of early-type stars.  
This publication makes use of data products from the Two Micron All Sky Survey, which is a joint project of the University of Massachusetts and the Infrared Processing and Analysis Center/California Institute of Technology, funded by the National Aeronautics and Space Administration and the National Science Foundation. IRAF is distributed by the National Optical Astronomy Observatory, which is operated by the Association of Universities for Research in Astronomy (AURA) under a cooperative agreement with the National Science Foundation. This work is based [in part] on observations made with the Spitzer Space Telescope, which is operated by the Jet Propulsion Laboratory, California Institute of Technology under a contract with NASA.
This research has made use of data obtained from the Chandra Source Catalog, provided by the Chandra X-ray Center (CXC) as part of the Chandra Data Archive. 
This research has made use of the VizieR catalogue access tool, CDS, Strasbourg, France. The original description of the VizieR service was published in A\&AS 143, 23


%% in parentheses, after the keyword, but they will not be verified.

{\it Facilities:} \facility{SOAR Telescope}, \facility{OSIRIS}, \facility{Goodman}.

\clearpage

\end{document}